\documentclass{emulateapj}
\usepackage{epsfig, graphicx} 
\usepackage{natbib}

\def\etal   {{\rm et\ts al.}}
\def\kms    {\ifmmode{{\rm \ts km\ts s}^{-1}}\else{\ts km\ts s$^{-1}$}\fi}
\def\msol   {\ifmmode{{\rm M}_{\odot}}\else{M$_{\odot}$}\fi}
\def\ts     {\thinspace} 
\def\ci   {\ifmmode{{\rm C}{\rm \small I}}\else{C\ts {\scriptsize I}}\fi}
\def\conel {\ifmmode{{\rm C}{\rm \small I}\,(^3P_1\to^3P_0)}\else{C\ts{\scriptsize I}\,{\small$(^3P_1\to^3P_0)$}}\fi}
\def\ctwol {\ifmmode{{\rm C}{\rm \small I}\,(^3P_2\to^3P_1)}\else{C\ts{\scriptsize I}\,{\small$(^3P_2\to^3P_1)$}}\fi}
\def\cone {\ifmmode{{\rm C}{\rm \small I}(1-0)}\else{C\ts {\scriptsize I}(1--0)}\fi}
\def\ctwo {\ifmmode{{\rm C}{\rm \small I}(2-1)}\else{C\ts {\scriptsize I}(2--1)}\fi}
\def\cii  {\ifmmode{{\rm C}{\rm \small II}}\else{C\ts {\scriptsize II}}\fi}
\def\aco  {\ifmmode{^{12}{\rm CO}(J=1\to0)}\else{$^{12}{\rm CO}(J=1\to0)$}\fi}
\def\bco  {\ifmmode{^{12}{\rm CO}(J=2\to1)}\else{$^{12}{\rm CO}(J=2\to1)$}\fi}
\def\m    {\ifmmode{\mu {\rm m}}\else{$\mu$m}\fi}
\def\cco  {\ifmmode{^{13}{\rm CO}(J=1\to0)}\else{$^{13}{\rm CO}(J=1\to0)$}\fi}
\def\dco  {\ifmmode{^{13}{\rm CO}(J=2\to1)}\else{$^{13}{\rm CO}(J=2\to1)$}\fi}
\def\eco  {\ifmmode{^{12}{\rm CO}(J=3-2)}\else{$^{12}{\rm CO}(J=3-2)$}\fi}
\def\hi   {\ifmmode{{\rm H}{\rm \small I}}\else{H\ts {\scriptsize I}}\fi}
\def\hii  {\ifmmode{{\rm H}{\rm \small II}}\else{H\ts {\scriptsize II}}\fi}
\def\ha   {\ifmmode{{\rm H}{\alpha}}\else{H${\alpha}$}\fi}
\def\hh     {\ifmmode{{\rm H}_2}\else{H$_2$}\fi}
\def\nhh     {\ifmmode{N({\rm H}_2)}\else{$N$(H$_2$)}\fi}
\def\tex {\ifmmode{{T}_{\rm ex}}\else{$T_{\rm ex}$}\fi}
\def\tmb {\ifmmode{{T}_{\rm mb}}\else{$T_{\rm mb}$}\fi}
\def\tkin {\ifmmode{{T}_{\rm kin}}\else{$T_{\rm kin}$}\fi}
\def\microns {\ifmmode{\mu{\rm m}}\else{$\mu$m}\fi}
\def\nhh   {\ifmmode{n({\rm H}_2)}\else{$n$(H$_2$)}\fi}

\shorttitle{Atomic Carbon at High Redshift}

\shortauthors{Walter, Wei{\ss} et al.}

\journalinfo{Submitted to ApJ}

\begin{document}

\title{A Survey of Atomic Carbon at High Redshift}
   \author{
          F. Walter
          \altaffilmark{1},
          A. Wei{\ss}
          \altaffilmark{2},
          D. Downes
          \altaffilmark{3},
          R. Decarli
          \altaffilmark{1},
          C. Henkel
          \altaffilmark{2}
           }
\altaffiltext{1}{Max-Planck-Institut f\"ur Astronomie, K\"onigstuhl 17, D-69117 Heidelberg, Germany    [e-mail: {\em walter@mpia.de}]}
\altaffiltext{2}{Max-Planck-Insitut f\"ur Radioastronomie, Auf dem H\"ugel 69, D-53121 Bonn, Germany}
\altaffiltext{3}{Institut de Radio Astronomie Millim\'etrique (IRAM), St. Martin d'H\`eres, France}

\begin{abstract} 

We present a survey of atomic carbon (\ci) emission in high--redshift
(z$>$2) submillimeter galaxies (SMGs) and quasar host galaxies
(QSOs). Sensitive observations of the \conel\ and \ctwol\ lines have
been obtained at the IRAM Plateau de Bure interferometer and the IRAM
30\,m telescope. A total of 16 \ci\ lines have been targeted in 10
sources, leading to a total of 10 detected lines --- this doubles the
number of \ci\ observations at high redshift to date. We include
previously published \ci\ observations (an additional 5 detected
sources) in our analysis.  Our main finding is that the \ci\
properties of high--redshift systems do not differ significantly from
what is found in low--redshift systems, including the Milky Way. The
\ctwol/\conel\ and the \conel/$^{12}$CO(3--2) line luminosity ($L'$)
ratios change little in our sample, with respective ratios of
0.55$\pm$0.15 and 0.32$\pm$0.13. The \ci\ lines are not an important
contributor to cooling of the molecular gas (average $L_\ci$/$L_{\rm
FIR}\sim (7.7\pm4.6)\times10^{-6}$). We derive a mean carbon
excitation temperature of 29.1$\pm$6.3\,K, broadly consistent with
dust temperatures derived for high--redshift starforming systems, but
lower than gas temperatures typically derived for starbursts in the
local universe. The carbon abundance of
X[\ci]/X[H$_2$]$\sim$(8.4$\pm3.5)\times$10$^{-5}$ is of the same order
as found in the Milky Way and nearby galaxies. This implies that the
high--z galaxies studied here are significantly enriched in carbon on
galactic scales, even though the look--back times are considerable
(the average redshift of the sample sources corresponds to an age of
the universe of $\sim$2\,Gyr).

\end{abstract}


\section{Introduction}

Detections of carbon monoxide (CO) and dust, in objects at $z$\,$>$2
have enabled studies of the molecular gas properties in the early
phases of galaxy formation (e.g., review by Solomon \& Vanden Bout
2005). Molecular gas masses $>10^{10}$\,\msol, enriched in C and O,
provide strong evidence for powerful starbursts at early epochs in
objects that may have later turned into massive spirals and/or
elliptical galaxies.  Until now, mid--$J$ ($J$=3--6) CO lines and the
continuum emission from dust have primarily been used to characterize
the physical properties of these massive gas reservoirs. One key
approach is to observe the different rotational ($J$) transitions of
CO (so--called `CO line spectral energy distribution', or `CO line
ladder') and use radiative transfer models to constrain molecular gas
masses, densities, and temperatures (see the compilation by Wei{\ss}
et al.\ 2007, 2011, hereafter W11). In such analyses, however, it is
frequently found that the density and kinetic temperature of the
molecular gas are degenerate, so that data from additional tracers of
the gas phase would be highly desirable.

Such an alternative tracer of the cold molecular gas phase is atomic
carbon (\ci). Studies of atomic carbon in the local universe have been
carried out in molecular clouds of the galactic disk, the galactic
center, M82 and other nearby galaxies (e.g., White et al.\ 1994;
Stutzki et al.\ 1997; Gerin \& Phillips 1998; 2000; Ojha et al.\ 2001;
Israel \& Baas 2002; Schneider et al.\ 2003, Israel 2005). These
studies have shown that \ci\, is closely associated with the CO
emission independent of environment.  Since the critical density for
the \cone\, and \aco\, lines are both $n_{\rm cr} \approx 10^3\,{\rm
cm}^{-3}$, this finding suggests that the transitions arise from the
same volume and share similar excitation temperatures (e.g. Ikeda et
al.\ 2002). For reference, the critical densities for the \cone\ and
\ctwo\ lines are 1$\times$10$^3$\,cm$^{-3}$ and
3$\times$10$^3$\,cm$^{-3}$, respectively (Sch{\"o}ier et al.\
2005)\footnote{For details see:
http://www.strw.leidenuniv.nl/\~\,moldata/datafiles/catom.dat}. As
pointed out by Papadopoulos et al.\ (2004) such a close connection
between \ci\ and CO was not expected from early theoretical work
(Tielens \& Hollenbach 1985a, 1985b), where the \ci\ emission was
thought to emerge from a narrow \cii/\ci/CO transition zone in
molecular clouds.

Because the $^3$P fine-structure system of atomic carbon forms a
simple three-level system, detection of both optically thin carbon
lines, \conel\ (rest frequency: 492.161\,GHz, hereafter \cone) and
\ctwol\ (rest frequency: 809.344\,GHz, hereafter \ctwo), enables one
to derive the excitation temperature, neutral carbon column density
and mass independently of any other information (e.g., Ojha et al.\
2001, Wei\ss\ et al.\ 2003).  A combination of this method (using \ci)
with the aforementioned CO SEDs is particularly powerful as it
eliminates some of the degeneracy frequently found in CO radiative
transfer models. Therefore, assuming that \ci\ and CO trace similar
volumes on galactic scales, the carbon excitation temperature allows
us to pin down the CO excitation conditions, as the excitation
temperature of \ci\ and CO(1--0) are expected to be similar. 

Previous observations of \ci\ at high redshift (see references given
in Section~4 and the related Table) have shown that both the \cone\
and \ctwo\ lines are typically significantly fainter than the
respective CO lines at similar frequencies (partly because they are
typically optically thin). In this paper we report on observations of
atomic carbon towards 10 high--redshift sources (a total of 16 \ci\
lines) that encompass both submillimeter galaxies (SMGs) as well as
quasar host galaxies (QSOs). These observations roughly double the
number of high--z \ci\ line detections that have been presented in the
literature previously (in a dozen papers). In Section~2 we present the
observations that have been obtainted with the IRAM Plateau de Bure
Interferometer (hereafter: PdBI) as well as the IRAM 30\,m
telescope. In Section~3 we discuss the observations of the individual
sources. We compare the \ci\ properties (including sources from the
literature) in Section~4. Section 5 summarizes our findings.
Throughout this paper we use a $\Lambda$ cosmology with $H_{\rm 0} =
71$ \kms\,Mpc$^{-1}$, $\Omega_\Lambda=0.73$ and $\Omega_m=0.27$
(Spergel \etal\ 2003, 2007).

\begin{figure}
\centering
\includegraphics[width=8.5cm,angle=0]{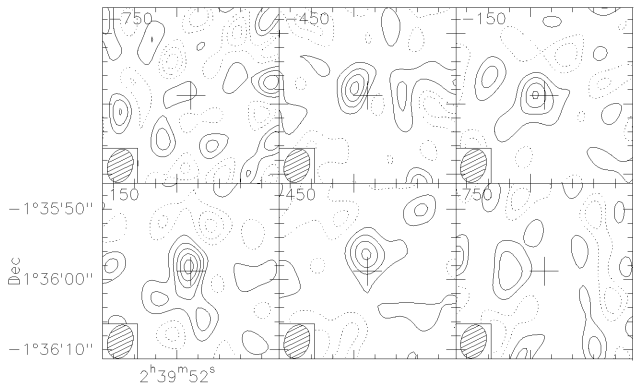}
\caption{\cone\ channel maps of SMM\,J02399--0136. Each channel is
300\kms\ wide and the central velocity is given in the
top left corner. Contours are shown in steps of 1$\sigma$
(1$\sigma$=0.31\,mJy\,beam$^{-1}$)}.
\end{figure}

\begin{figure*}
\centering
\includegraphics[width=13.0cm,angle=0]{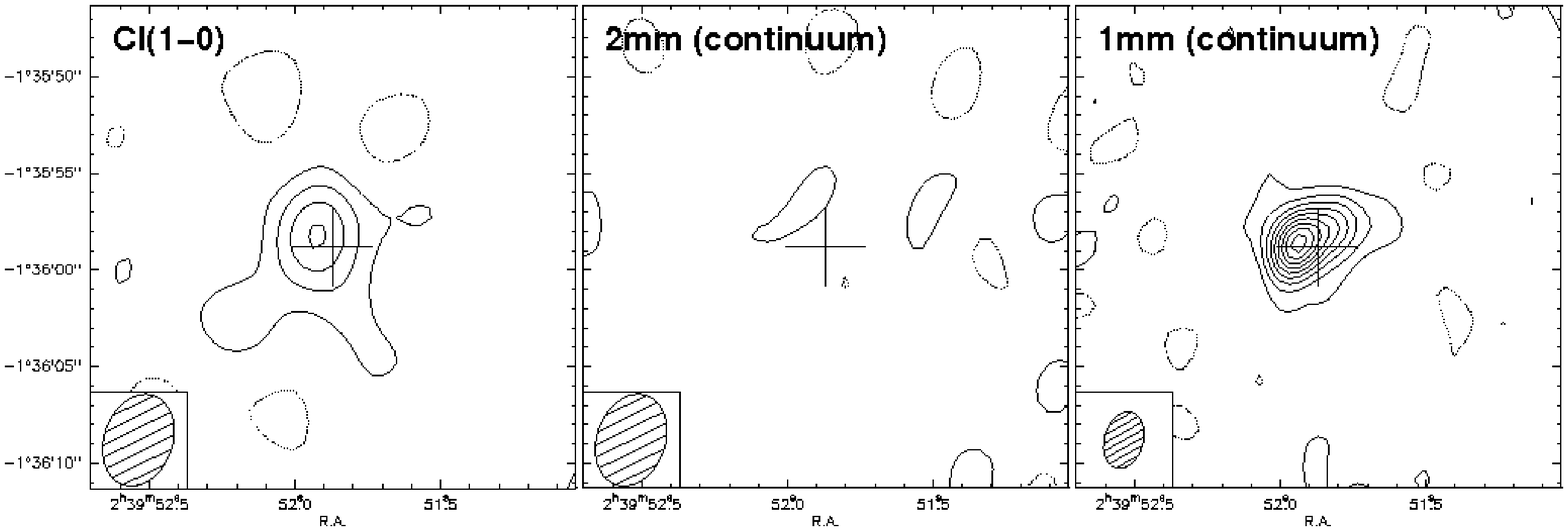}
\caption{Results for SMM\,J02399--0136. {\em Left:} Integrated \cone\
emission from --500\kms\ to +500\kms, with contours starting at
$\pm$2$\sigma$ in steps of 2$\sigma$ (1\,$\sigma$=0.17\,mJy). {\em
Middle:} Continuum measurement at 129.2\,GHz (line--free channels of the
\cone\ observations, using the same total bandwidth as for the line), same 
contouring and 1$\sigma$ sensitivity as in
the left panel. No continuum is detected at 129.2\,GHz. {\em Right:} 
Continuum  emission at 212.5\,GHz, with
contours starting at $\pm$2$\sigma$ in steps of 2$\sigma$
(1\,$\sigma$=0.18\,mJy). The  beam sizes are shown in the
bottom left corner, respectively (see Table~2).}
\end{figure*}

\begin{figure}
\centering
\includegraphics[width=8.5cm,angle=0]{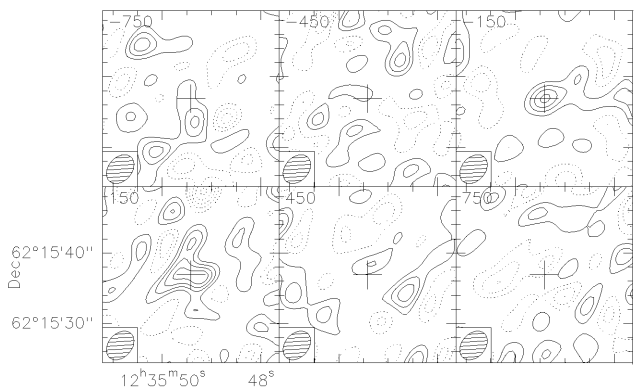}
\caption{\cone\ channel maps of SMM\,J123549+6215. Each channel is
300\kms\ wide and the central velocity is given in the
top left corner. Contours are shown in steps of 1$\sigma$
(1$\sigma$=0.43\,mJy\,beam$^{-1}$)}.
\end{figure}

\begin{figure*}
\centering
\includegraphics[width=13.0cm,angle=0]{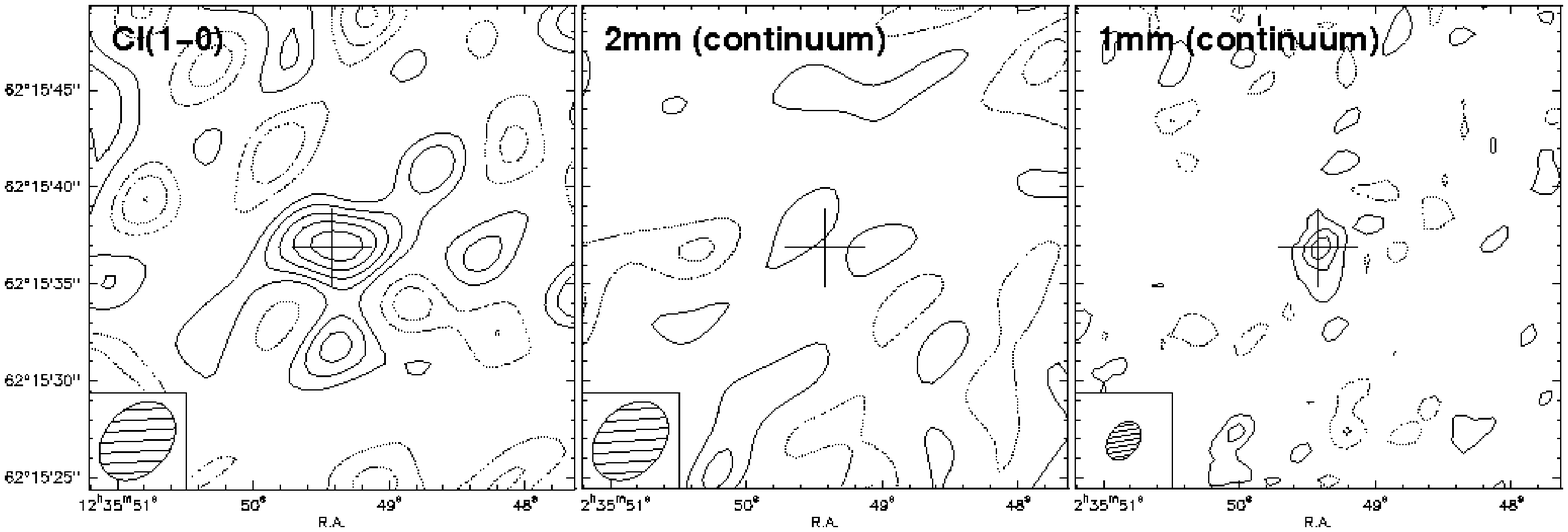}
\caption{Results for SMM\,J123549+6215. {\em Left:} Integrated \cone\
emission from --400\kms\ to +400\kms, with contours starting at
$\pm$1$\sigma$ in steps of 1$\sigma$ (1\,$\sigma$=0.26\,mJy). {\em
Middle:} Continuum measurement at 153.3\,GHz (line--free channels of the
\cone\ observations, using the same total bandwidth as for the line), same contouring and 1$\sigma$ sensitivity as in
the left panel. No continuum is detected at 153.3\,GHz. {\em Right:} 
Continuum emission at 252.8\,GHz, with
contours starting at $\pm$2$\sigma$ in steps of 2$\sigma$
(1\,$\sigma$=0.20\,mJy). The  beam sizes are shown in the
bottom left corner, respectively (see Table~2).}
\end{figure*}

\section{Observations}

\subsection{Source Selection} 

Given the expected faintness of the \ci\ emission line we have
selected sources that are bright in CO emission. We have mostly
targeted quasars using the IRAM 30\,m telescope as their linewidths
are typically small ($\sim400$\,km\,s$^{-1}$), to minimize
susceptibility to low--frequency baseline instablities in the single
dish observations. We have typically used the (intrinsically more
stable) PdBI in the case of the SMGs that are faint and have
significant linewidths ($\sim800-1000$\,km\,s$^{-1}$, e.g. Greve et
al.\ 2005).  The SMGs have been selected to be in the redshift range
2.03$<$z$<$2.81 so that both \ci\ transitions can be observed in the 2
and 1\,mm bands of the PdBI and 30\,m telescope. As discussed in the
individual sources below, some of the SMGs host an AGN, i.e.\ the two
populations that have historically been referred to as SMGs and
quasars do overlap. Most of the quasars in our sample have higher
redshifts.  The source names, coordinates, redshifts, targeted \ci\
transitions and telescopes used are given in Table~1 and results for
individual sources are discussed in Section~3.

\begin{deluxetable*}{llllllll}
\tablecaption{Properties of Observed Targets}
\tablewidth{0pt}
\tablehead{
\colhead{source} & \colhead{RA} & \colhead{DEC}  & \colhead{z} & \colhead{FWHM} & \colhead{transition} & \colhead{telescope} & \colhead{references\tablenotemark{a}}  \\
\colhead{}    & \colhead{J2000.0}  & \colhead{J2000.0}&  \colhead{}   &  \colhead{km\,s$^{-1}$}        &  \colhead{}            &  \colhead{} 
}
\startdata
SMM\,J02399--0136                 & 02:38:51.87  & --01:35:58.8 & 2.808  & $\sim$1100  &\cone, \ctwo & PdBI  & 1, 2   \\
APM\,08279+5255                   & 08:31:41.64  &  +52:45:17.5 & 3.911  & 480$\pm$35  &\ctwo        & 30\,m & 3      \\
RX\,J0911+0551                    & 09:11:27.50  &  +05:50:52.0 & 2.796  & 150$\pm$20  &\cone, \ctwo & 30\,m & 4, 5   \\
SMM\,J123549+6215\tablenotemark{b}& 12:35:49.42  &  +62:15:36.9 & 2.202  & 600$\pm$50  &\cone, \ctwo & PdBI  & 6      \\
BRI\,1335--0417                   & 13:38:03.42  & --04:32:34.1 & 4.407  & 420$\pm$60  & \ctwo       & 30\,m & 7      \\
SMM\,J14011+0252                  & 14:01:04.93  &  +02:52:24.1 & 2.565  & 190$\pm$11  &\ctwo        & 30\,m & 8, 9   \\
SMM\,J16359+6612                  & 16:35:54.10  &  +66:12:23.8 & 2.517  & 500$\pm$100 &\cone, \ctwo & 30\,m & 10, 11 \\
SMM\,J163650+4057\tablenotemark{c} & 16:36:50.41  &  +40:57:34.3 & 2.385  & 710$\pm$50  &\cone, \ctwo & PdBI  & 12, 6  \\
SMM\,J163658+4105\tablenotemark{d} & 16:36:58.17  &  +41:05:23.3 & 2.452  & 800$\pm$50  &\cone, \ctwo & PdBI  & 13, 6  \\
PSS\,J2322+1944                   & 23:22:07.25  &  +19:44:22.1 & 4.120  & 200$\pm$70  &\ctwo        & 30\,m & 14     
\enddata
\tablecomments{The coordinates and redshifts refer to the pointing center and redshift used for tuning the receivers. These values are taken from the references given in the last column.}
\tablenotetext{a}{references (coordinates, redshifts, FWHM's) are: 
[1]: Frayer \etal\ 1998; 
[2]: Genzel \etal\ 2003; 
[3]:  Downes \etal\ 1999; 
[4]: Hainline \etal\ 2004; 
[5]: W11; 
[6]: Tacconi \etal\ 2006; 
[7]: Guilloteau et al.\ 1999;
[8]: Frayer \etal\ 1999; 
[9]: Downes \& Solomon 2003; 
[10]: Sheth \etal\ 2004; 
[11]: Kneib \etal\ 2005, 
[12]: Neri \etal\ 2003; 
[13]: Greve \etal\ 2005; 
[14]: Cox \etal\ 2002.
}
\tablenotetext{b}{alternative name: HDF\,76.}
\tablenotetext{c}{alternative name: Elias N2 850.4.}
\tablenotetext{d}{alternative name: Elias N2 850.2.}
\end{deluxetable*}

\subsection{IRAM Plateau de Bure Observations}

The data were taken in the years 2008 and 2009 (project IDs S046 \&
S0B3) using the upgraded PdBI with its dual polarization receivers and
a bandwidth of 1\,GHz (single polarization). At 2\,mm, typically five
15m antennas were available in the compact D configuration, with
spacings from 24 to 86m.  These short spacings had average rms phase
errors of 15$^\circ$ at 2\,mm and 30$^\circ$ at 1\,mm.  System
temperatures were 150K in the lower sideband at 2\,mm, and 250 to 400K
in the lower sideband at 1.3mm. The spectral correlators covered
$\sim$1500\kms\ at 2\,mm and 1000\kms\ at 1\,mm, with instrumental
resolutions of 8 and 4\kms, respectively. The primary amplitude
calibrators were 3C454.3 (variable, but typically 25 Jy at 2\,mm and
13 Jy at 1.3mm), and MWC349 (nonvarying, with 1.9 and 1.6 Jy at 2 and
1.3mm). The typical uncertainties in the flux scales and overall
calibration are $\sim$10\% at 2\,mm and $\sim$15--20\% at 1.3mm and
should be added to the formal fitting errors quoted in the remainder
of this paper. The observing program monitored phases every
20\,min. The data processing program used water-vapour monitoring
receivers at 22 GHz on each antenna to correct amplitudes and phases
at 2\,mm and 1.3\,mm for short--term changes in atmospheric water
vapor. All visibilities are weighted by the integration time and the
inverse square of the system temperature. Source maps were made with
`natural weighting' of the visibilities to maximise the
sensitivity. The tuned frequencies, the resulting beam sizes (and
position angles), and the 1$\sigma$ rms over the full bandwidth
(1\,GHz) for each source are given in Table~2. The primary beam size
of PdBI at 2\,mm (1\,mm) is $\sim$35$''$ ($\sim$22$''$). Emission has
been cleaned to the 1.5\,$\sigma$ level in a tight box around the
source. In the case of non--detections no cleaning was performed.

\subsection{IRAM 30\,m Observations}

Observations were carried out with the IRAM 30\,m telescope in the
years 2004--2007. We used the CD or AB receiver configurations with
the C/D 150 receivers tuned to the \cone\ transition or the A/B 230
receivers tuned to \ctwo. The beam size of the 30\,m at 140\,GHz
(230\,GHz) is $\approx\,17''$ (11$''$).  Typical system temperatures
were $\approx$\,200\,K and $\approx$\,270\,K during winter and summer
respectively. The observations were carried out in wobbler switching
mode, with a switching frequency of 0.5\,Hz and a wobbler throw of
$50''$ in azimuth.  Pointing was checked frequently and was found to
be stable within $3''$. Calibration was obtained every 12\,min using
standard hot/cold--load absorber measurements. Both Mars and Uranus
were used for calibration. The antenna gain was found to be consistent
with the standard value of 6.6 (10) Jy\,K$^{-1}$ at 140\,(230)\,GHz
(temperatures in units of T$_{\rm A}^*$). We estimate the flux density
scale to be accurate to about 15\%.

Data were recorded using the 4\,MHz filter banks on each receiver (512
channels, 1GHz bandwidth, 4\,MHz channel spacing) and were processed
using the CLASS software. After dropping bad data only linear
baselines were subtracted from individual spectra. The tuned
frequencies and the noise in each 150\kms\ channel are given in
Table~3.

\begin{figure*} \centering
\includegraphics[width=13.0cm,angle=0]{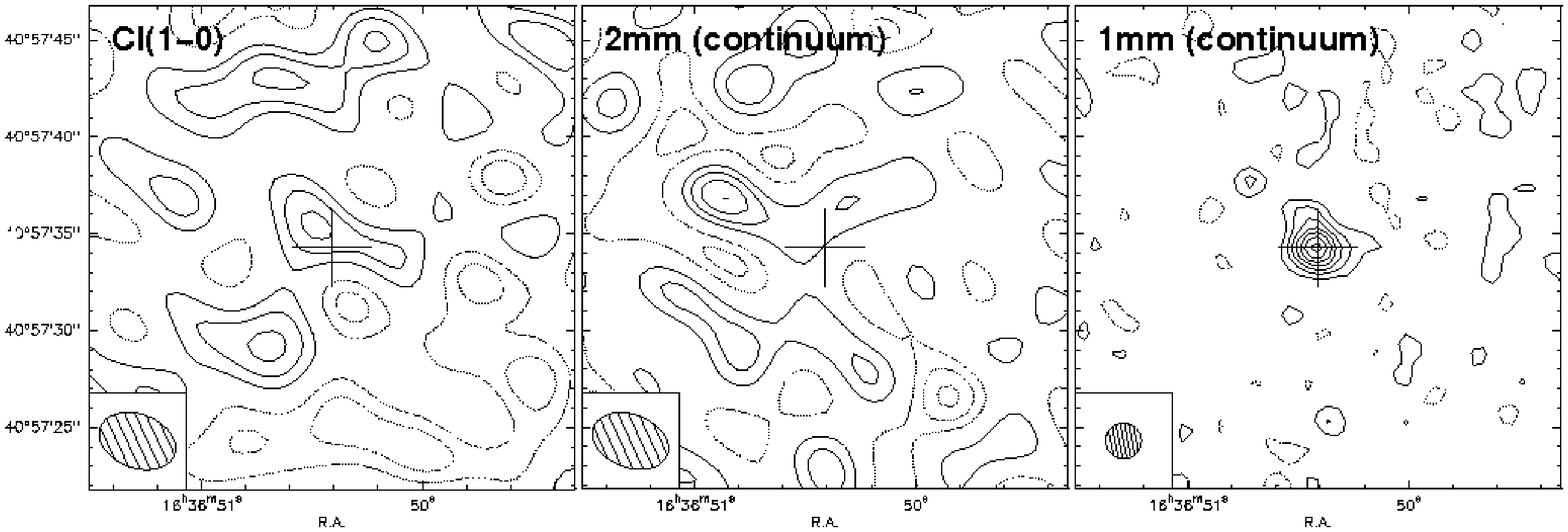}
\caption{Results for SMM\,J163650+4057. {\em Left:} Integrated \cone\
emission from --500\kms\ to +500\kms, with contours starting at
$\pm$1$\sigma$ in steps of 1$\sigma$ (1\,$\sigma$=0.24\,mJy). The
\cone\ line is not detected. {\em Middle:} Continuum measurement at
145.4\,GHz (line--free channels of the \cone\ observations, using the
same total bandwidth as for the line), same contouring and 1$\sigma$
sensitivity as in the left panel. The continuum at 145.4\,GHz is not
detected. {\em Right:} Continuum emission at 239.1\,GHz, with contours
starting at $\pm$2$\sigma$ in steps of 2$\sigma$
(1\,$\sigma$=0.22\,mJy). The beam sizes are shown in the bottom left
corner, respectively (see Table~2).}  \end{figure*}

\begin{figure*}
\centering
\includegraphics[width=13.0cm,angle=0]{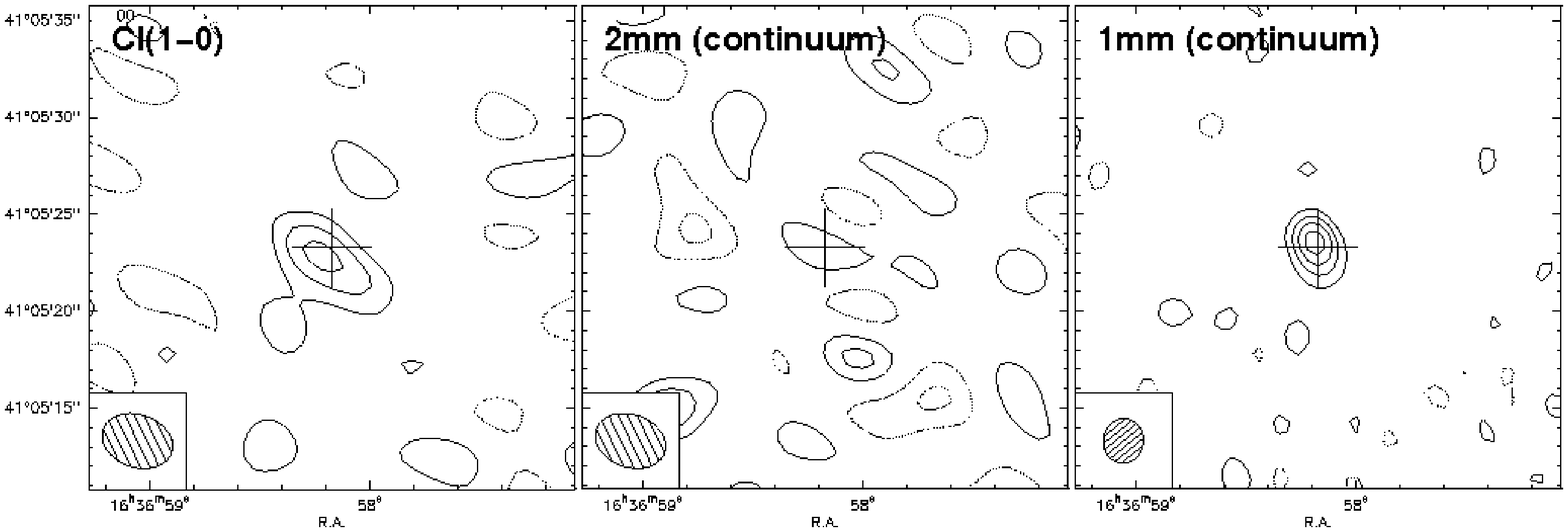}
\caption{Results for SMM\,J163658+4105. {\em Left:} Integrated \cone\
emission from --400\kms\ to +200\kms, with contours starting at
$\pm$1.5$\sigma$ in steps of 1.5$\sigma$ (1\,$\sigma$=0.29\,mJy). {\em
Middle:} Continuum measurement at 142.6\,GHz (line--free channels of the
\cone\ observations, using the same total bandwidth as for the line), same 
contouring and 1$\sigma$ sensitivity as in
the left panel. There is a tentative detection of the 142.6\,GHz continuum
in this source. {\em Right:} Continuum emission at 234.5\,GHz, with
contours starting at $\pm$2$\sigma$ in steps of 2$\sigma$
(1\,$\sigma$=0.22\,mJy). The beam sizes are shown in the
bottom left corner, respectively (see Table~2).}
\end{figure*}

\begin{deluxetable*}{lllllllll}
\tablecaption{Plateau de Bure Observations}
\tablewidth{0pt}
\tablehead{
\colhead{source} & 
\colhead{FWZI\tablenotemark{a}} & 
\ci\ trans. & 
\colhead{$\nu_{\rm obs}$}  & 
\colhead{beamsize, P.A.}&
\colhead{rms\tablenotemark{b}} & 
\colhead{I$_{\rm CI}$\tablenotemark{c}} & 
\colhead{S$_{\rm cont}$\tablenotemark{d}} & 
\colhead{S$_{\rm exp}$\tablenotemark{e}} \\
\colhead{} &  
\colhead{[km\,s$^{-1}$]}  & 
&
\colhead{[GHz]}  & 
\colhead{$"\times"$, $^\circ$} &
\colhead{[mJy\,b$^{-1}$]} & 
\colhead{[Jy\,km\,s$^{-1}$]} & 
\colhead{[mJy]} & 
\colhead{[mJy]} \\
}
\startdata
SMM\,02399--0136 & $\sim$1000 & \cone & 129.2439 &    4.9$\times$3.5, 161 & 0.11 & 1.9$\pm$0.2		 & $<$0.5			& 1.2$\pm$0.5  \\
               &            & \ctwo & 212.5373 &    3.0$\times$2.0, 165 & 0.18 & 1.8$\pm$0.9		 & 6.2$\pm$0.2\tablenotemark{i}      &   7$\pm$3    \\     
SMM\,123549+6215 & $\sim$800  & \cone & 153.7041 &    4.6$\times$3.3, 139 & 0.17 & 1.1$\pm$0.2		 & $<$0.8			& 0.8$\pm$0.3  \\
	       &            & \ctwo & 252.762  &    2.2$\times$1.5, 145 & 0.20 & $<$1.0 		 & 1.7$\pm$0.2  		& 3.4$\pm$1.3  \\
SMM\,163650+4057  & $\sim$1400 & \cone & 145.3946 &    4.1$\times$2.8, 69  & 0.17 & $<0.7$\tablenotemark{f} & $<$0.8			& 0.7$\pm$0.3  \\
 	       &            & \ctwo & 239.0966 &    1.9$\times$1.8, 99  & 0.18 & $<$2.0 		 & 3.7$\pm$0.2\tablenotemark{g} & 2.9$\pm$1.2  \\
SMM\,163658+4105  & $\sim$1000 & \cone & 142.5726 &    3.7$\times$2.8, 69  & 0.16 & 0.9$\pm$0.2		 & $\sim$0.6\tablenotemark{h}	& 0.8$\pm$0.3  \\
               &            & \ctwo & 234.456  &    3.6$\times$1.4, 174 & 0.21 & $<$2.0 		 & 2.0$\pm$0.2  		& 3.3$\pm$0.6 
\enddata
\tablecomments{The uncertainties given here are only based on the formal fitting error and do not include the uncertainties in flux calibration (as discussed in Sec.~2.2).}
\tablenotetext{a}{The full width zero intensity (FWZI) has been estimated from published CO spectra (see references in Table~1).}
\tablenotetext{b}{Rms in mJy\,beam$^{-1}$ over the full 1\,GHz bandwidth. }
\tablenotetext{c}{Upper limits to the \ctwo\ line fluxes are 5$\sigma$ values (see Section~3).}
\tablenotetext{d}{Measured continuum flux density at frequency given in column 4. In the case of non--detections we quote 3$\sigma$ upper limits.}
\tablenotetext{e}{Expected continuum at observed frequency based on SCUBA 850 $\mu$m measurements (Table~4, Column 7) and assuming a modified blackbody with $\beta$=1.5 and a dust temperature of T=34.6\,K (Kov{\'a}cs et al.\ 2006). The uncertainties reflect the model uncertainties of 40\%, as discussed in Section~4.2.2.}
\tablenotetext{f}{Emission is marginally detected in the blue wing of the CO(3--2) line (Neri et al.\ 2003).}
\tablenotetext{g}{This source is potentially resolved in the continuum. The peak flux density is  2.8$\pm$0.2\,mJy.}
\tablenotetext{h}{This source is is only marginally detected in the 2\,mm continuum ($<$3$\sigma$).
\tablenotetext{i}{This flux likely includes the \ctwo\ line, see Sec.~3.1.3.}
}
\end{deluxetable*}

\newpage
\section{Results}

In the following we present the results for each source. The PdBI
observations are summarized in Section~3.1 (Figs.~1--6) and the 30\,m
observations in Section~3.2 (Figs.~7--12). 

\subsection{PdBI Observations}

\subsubsection{\ci\ Line Emission}

The main results and data analysis are described in the following:
\cone\ emission (observed at 2\,mm wavelength) has been detected at a
significant level in three of the four sources that have been
targeted. However, \ctwo\ emission (observed at 1\,mm wavelengths) has
been tentatively detected only in one source (SMM\,J02399--0136). The
latter measurement is complicated by the fact that the faint \ctwo\
emission is situated on top of bright continuum emission at 1\,mm
wavelengths. An additional complication is that the rest frequency of
the CO(7--6) line (806.651\,GHz) is close to that of \ctwo\
(809.344\,GHz), i.e.\ the velocity difference at z$\sim$2.5 is
($\Delta \nu$)/$\nu\,\times$ c$\sim$1000\kms\ (where c is the speed of
light). This means that the lines begin to overlap in frequency space
at this redshift if the intrinsic full width zero intensity (FWZI) is
larger than 1000\kms. This is not the case in most of our sources, but
potentially complicates the determination of the continuum redward of
the \ctwo\ line. In the light of this we present conservative
5$\sigma$ upper limits for the \ctwo\ line fluxes.  For the 2\,mm
continuum we give 3$\sigma$ upper limits. The $\sigma$'s used for
deriving the upper limits are based on the 1$\sigma$ noise in a
channel that has the expected FWHM of the respective line.

\begin{deluxetable*}{lllllll}
\tablecaption{IRAM 30\,m Observations}
\tablewidth{0pt}
\tablehead{
\colhead{source}  & \ci\ trans. & \colhead{$\nu_{\rm obs}$} & \colhead{rms\tablenotemark{a}} & \colhead{F$_{\rm peak}$} & \colhead{FWHM} & \colhead{I$_{\rm CI}$}\\
\colhead{}  & & \colhead{[GHz]} & \colhead{[mJy\,beam$^{-1}$]} & \colhead{[mJy]} & \colhead{[km\,s$^{-1}$]} & \colhead{[Jy\,km\,s$^{-1}$]}
}
\startdata
APM\,08279+5255    &     \ctwo &  164.7988  & 1.6 (85\,kms)& $<$4.8  & ---                       & $<$1.1\tablenotemark{a} \\
RX\,J0911+0551     &     \cone &  129.6524  & 2.9 (40\,kms)& 13.3    & $140\pm25$                & 2.1$\pm$0.3             \\
                   &     \ctwo &  213.2096  & 3.8 (45\,kms)& 14.8    & $145\pm30$                & 2.3$\pm$0.4             \\
BRI\,1335--0417    &     \cone &   91.0229  & 4.8 (60\,kms)& $<$14.6 & ---                       & $<$2.2\tablenotemark{b} \\
                   &     \ctwo &  149.6841  & 1.7 (60\,kms)& $<$5.1  & ---                       & $<$0.8\tablenotemark{b} \\
SMM\,J14011+0252   &     \ctwo &  227.0121  & 2.3 (50\,kms)& 14.4    & $205\pm25$                & 3.1$\pm$0.3		  \\
SMM\,J16359+6612   &     \cone &  139.9229  & 0.8 (75\,kms)& 3.9,3.8 & 250, 160\tablenotemark{c} & 1.7$\pm$0.3		  \\
	           &     \ctwo &  230.0991  & 0.9 (80\,kms)& 5.4,4.4 & 250, 160\tablenotemark{c} & 2.2$\pm$0.3		  \\
PSS\,J2322+1944    &     \ctwo &  158.0780  & 3.2 (60\,kms)& 8.0     & $160\pm55$                & 1.4$\pm$0.3		  
\enddata
\tablenotetext{a}{Velocity resolution for given rms shown in brackets.}
\tablenotetext{b}{3$\sigma$ upper limit over the full width of the expected line width.}
\tablenotetext{c}{Profile fit with two Gaussian fits (e.g. Wei{\ss} et al.\ 2005a).}
\end{deluxetable*}

\subsubsection{Continuum Emission} 

Continuum emission at 1\,mm has been detected in all cases (only one
object is tentatively detected in the 2\,mm band). We estimated the
expected 1\,mm and 2\,mm continuum flux densities (for the respective
frequency) based on published SCUBA (850\,$\mu$m) measurements
of the indvidual sources (column 7 in Table~4), assuming a modified
black body (see more detailed discussion in Section~4.2.2)\footnote{We
derive almost identical estimates using an Arp\,220 spectral energy
distribution}. We use the best--fit values for SMGs by Kov{\'a}cs et
al.\ (2006, T$_{\rm dust, SMG}$=34.6$\pm$3\,K, $\beta_{\rm SMG}$=1.5)
and assign uncertainties of $\pm10$\,K to T$_{\rm dust}$ and $\pm$0.2
to $\beta$.  In the case of the 1\,mm observations, the measured flux
densities of all our targets are consistent with these extrapolations
and inspection of individual datacubes did not reveal the presence of
an additional \ctwo\ line. No significant continuum emission was
detected in the 2\,mm data, consistent with expectations (Table~2).

All derived line fluxes and continuum flux densities are summarized in
Table~2. In the following we briefly discuss the individual sources.

\subsubsection{SMM\,J02399--0136}

This source is one of the first SMGs (S$_{850\mu m}$=26$\pm$3\,mJy,
Ivison et al.\ 1998) ever detected in CO emission ($J$=3, Frayer et al.\
1998)\footnote{$J$=3: CO(3--2)}. It has subsequently been studied in
detail in CO emission ($J$=3: Genzel et al.\ 2003, $J$=1: Ivison et al.\
2010a). Ivison et al.\ (1998) noted that this source hosts a dusty AGN;
its QSO lines are discussed in Villar-Mart{\'{\i}}n et al.\ (1999) and
Vernet \& Cimatti (2001). In the $L_{\rm FIR}$--$L_{\rm X-ray}$ plane
SMM\,J02399--0136 has the same luminosities as the Cloverleaf
(Alexander et al.\ 2005a). Ivison et al.\ (2010b) conclude that this
source comprises a merger between a FIR--luminous starburst, a QSO
host and a faint third component.

For our analysis we adopt a FWZI of $\sim$1000\kms\ for
SMM\,J02399--0136. The \cone\ line is clearly detected in both
individual channel maps (Figure~1) and the integrated line emission
(Figure~2, left). The total flux of the \cone\ line is
1.9$\pm$0.2\,Jy\,\kms. The continuum at 2\,mm is not detected at a
3$\sigma$ limit of S$_{\rm 129\,GHz}<$0.51\,mJy (Figure~2,
middle). The source is clearly resolved in the 1\,mm continuum at
212.5\,GHz with a total flux density of S$_{\rm
212\,GHz}$=6.2$\pm$0.2\,mJy (Figure~2, right). Given the large
linewidths in SMM\,J02399--0136 it is difficult to separate the
continuum from possible \ctwo\ emission in this source at 1\,mm
wavelengths. Genzel et al.\ (2003) have derived a total continuum flux
of 7.0$\pm$1.2\,mJy measured at 235\,GHz. If we assume that this is
the correct continuum flux, we would expect a 212\,GHz flux of
$\sim$4.9\,mJy, or about 1.3\,mJy less than what is observed. We
conclude that some of the measured flux may be attributed to \ctwo\
emission in this source and derive a tentative \ctwo\ flux of
$\sim$1.8\,Jy\kms\ with a considerable error bar (50\%). We note that
SMM\,J02399--0136 is thought to be lensed (magnification factor
$\mu$=2.5, Frayer et al.\ 1998, Genzel et al.\ 2003).

\subsubsection{SMM\,J123549+6215}

This source is sometimes also referred to as HDF\,76 (S$_{850\mu
m}$=8.3$\pm$2.5\,mJy, Chapman et al.\ 2003) and has been detected in
CO emission ($J$=3 and $J$=6: Tacconi et al.\ 2006, 2008, $J$=1:
Ivison et al.\ 2010a).  SMM\,J123549+6215 is X--ray bright (Alexander et
al.\ 2005a) and its luminosity is classified as AGN--dominated (see
also Alexander et al.\ 2005b, Takata et al.\ 2006).

We adopt a FWZI of $\sim$800\kms\ for SMM\,J123549+6215. The \cone\
line is detected both in individual channel maps (Figure~3) and in the
integrated line emission (Figure~4, left) with a total flux of
1.1$\pm$0.21\,Jy\,\kms. The continuum at 2\,mm is not detected at a
3$\sigma$ flux density limit of S$_{\rm 153\,GHz}<$0.78\,mJy
(Figure~4, middle). The 1\,mm continuum emission is detected at
S$_{\rm 252.8\,GHz}$=1.7$\pm$0.2 mJy (Figure~4, right). This is only
marginally consistent with the tentative detection quoted by Tacconi
et al.\ 2006 (S$_{\rm 213\,GHz}$=2.0$\pm$0.6\,mJy, the extrapolated
flux at 213\,GHz would be $\sim$1\,mJy), and within the expectations
based on the extrapolation from the SCUBA measurement (Table~2). We
derive a 5$\sigma$ upper limit of the \ctwo\ line (assuming a
bandwidth of 800\kms) of $<$1.0\,Jy\kms.

\subsubsection{SMM\,J163650+4057}

This source is also known as Elias N2\,850.4 (S$_{850\mu
m}$=8.2$\pm$1.7\,mJy, Ivison et al.\ 2002) and first CO data ($J$=3
and $J$=7) have been presented by Neri et al.\ (2003)\footnote{In Neri
et al.\ (2003) this source is referred to as SMM\,J16368+4057} ($J$=1
data are shown in Ivison et al.\ 2010a). Improved CO data are shown in
Tacconi et al.\ (2006, 2008, $J$=3 and $J$=6). The AGN broad lines of
this source (which in fact is a pair of blue and red galaxies, Ivison
et al.\ 2002) are discussed in Swinbank et al.\ (2005, see also Takata
et al.\ 2006, Men{\'e}ndez-Delmestre 2007, 2009).

The \cone\ line is not detected in our data with a 3$\sigma$ upper
limit across the expected linewidth ($\sim$1000\kms) of 0.7\,Jy\,\kms\
(Figure~5, left, but see caption of Table 2).  Likewise, the continuum
at 2\,mm is not detected at a 3$\sigma$ flux density limit of S$_{\rm
145\,GHz}<$0.74\,mJy (Figure~5, middle). The 1\,mm continuum emission
appears to be slightly extended with S$_{\rm 239.1\,GHz}$=3.7$\pm$0.4
mJy (peak flux: 2.8$\pm$0.18\,mJy, Figure~5, right), in rough
agreement with our SCUBA extrapolation and with Neri et al.\ (2003)
and Tacconi et al.\ (2006) who quote consistent 1\,mm flux densities
for this source (S$_{\rm 230\,GHz}$=2.5$\pm$0.4\,mJy and
2.6$\pm$0.5\,mJy, respectively). We derive a 5$\sigma$ upper limit of
the \ctwo\ line (assuming a bandwidth of 1000\kms) of $<$2.0\,Jy\kms.

\subsubsection{SMM\,J163658+4105}

This source is also known as Elias N2\,850.2 (S$_{850\mu
m}$=10.7$\pm$2.0\,mJy, Ivison et al.\ 2002) and CO data are published
in Greve et al.\ (2005, $J$=3)\footnote{In Greve et al.\ (2005) the
source is referred to as SMM\,J16366+4105}, Tacconi et al.\ (2006,
$J$=3 and $J$=7) and Ivison et al.\ 2010a ($J$=1). 

The \cone\ line is detected at 0.90$\pm$0.18\,Jy\,\kms\ over a
velocity width of 600\,\kms\ (Figure~6, left). This is the only source
in which the 2\,mm continuum emission may be marginally detected at
S$_{\rm 142.6\,GHz}\sim$0.6\,mJy. If real (as suggested by the good
match to the extrapolated SCUBA flux density, Table~2), the continuum
emission would contribute $\sim$40\% to the \cone\ flux density. Given
the low S/N of the potential continuum flux, we do not apply this
correction to the \cone\ flux in the analysis that follows but note
that such a correction would not change the statistical analysis
below. The 1\,mm continuum emission is detected at S$_{\rm
234\,GHz}$=2.0$\pm$0.22 mJy (Figure~6, right), in good agreement with
the extrapolation from the SCUBA measurement (and consistent with the
estimate of Tacconi et al.\, 2006, of S$_{\rm
233\,GHz}$=1.5$\pm$0.5\,mJy, their Table~1).

\begin{figure}
\begin{center}
\includegraphics[width=7cm,angle=0]{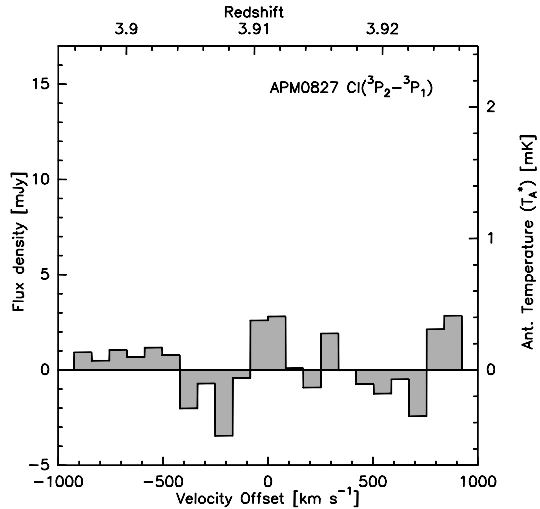}
\end{center}
\caption{{\em Left:} \ctwo\ spectrum of APM\,08279+5255; the line is not detected.}
\end{figure}

\begin{figure*}
\begin{center}
\includegraphics[width=7cm,angle=0]{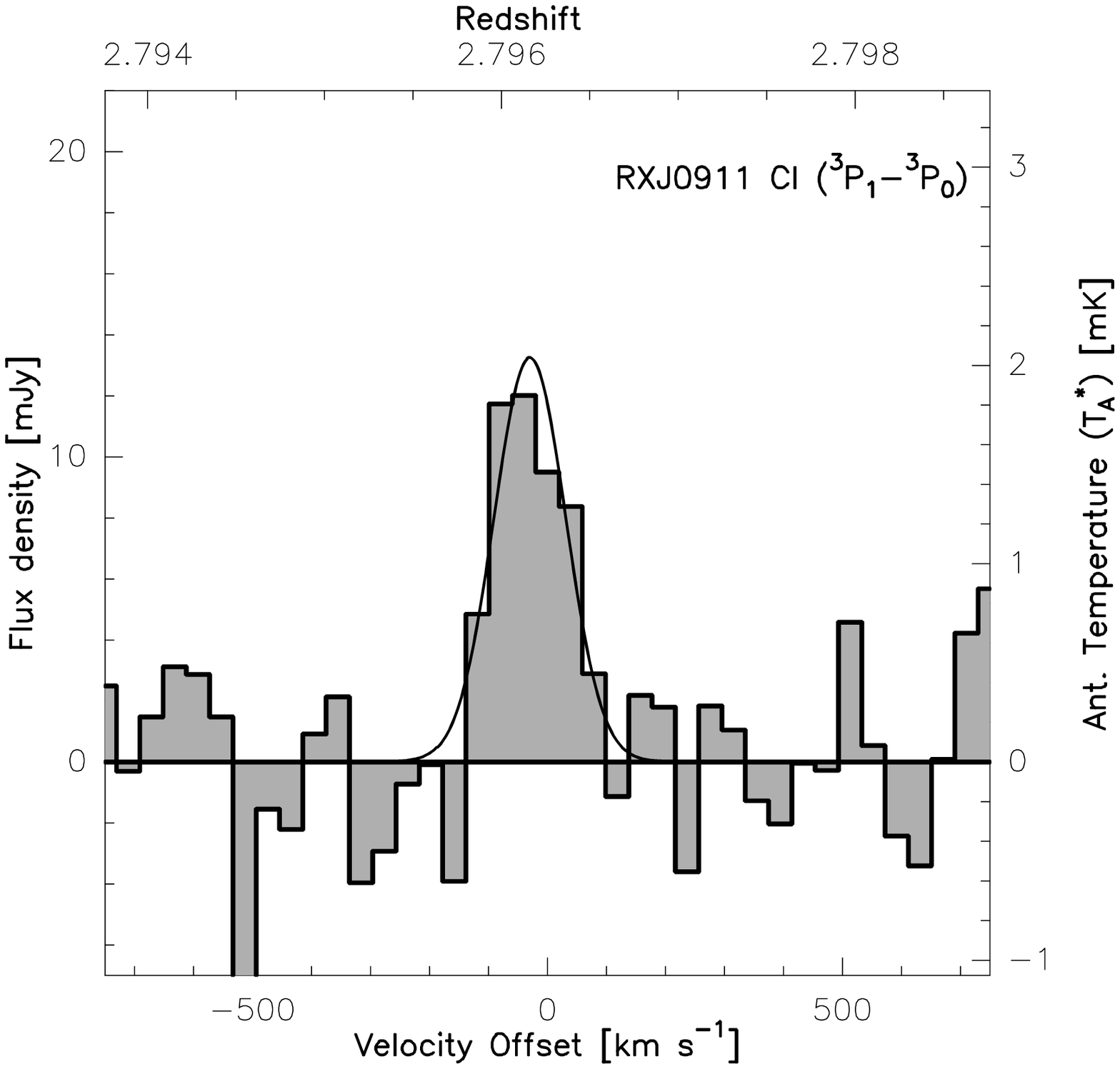}
\includegraphics[width=7cm,angle=0]{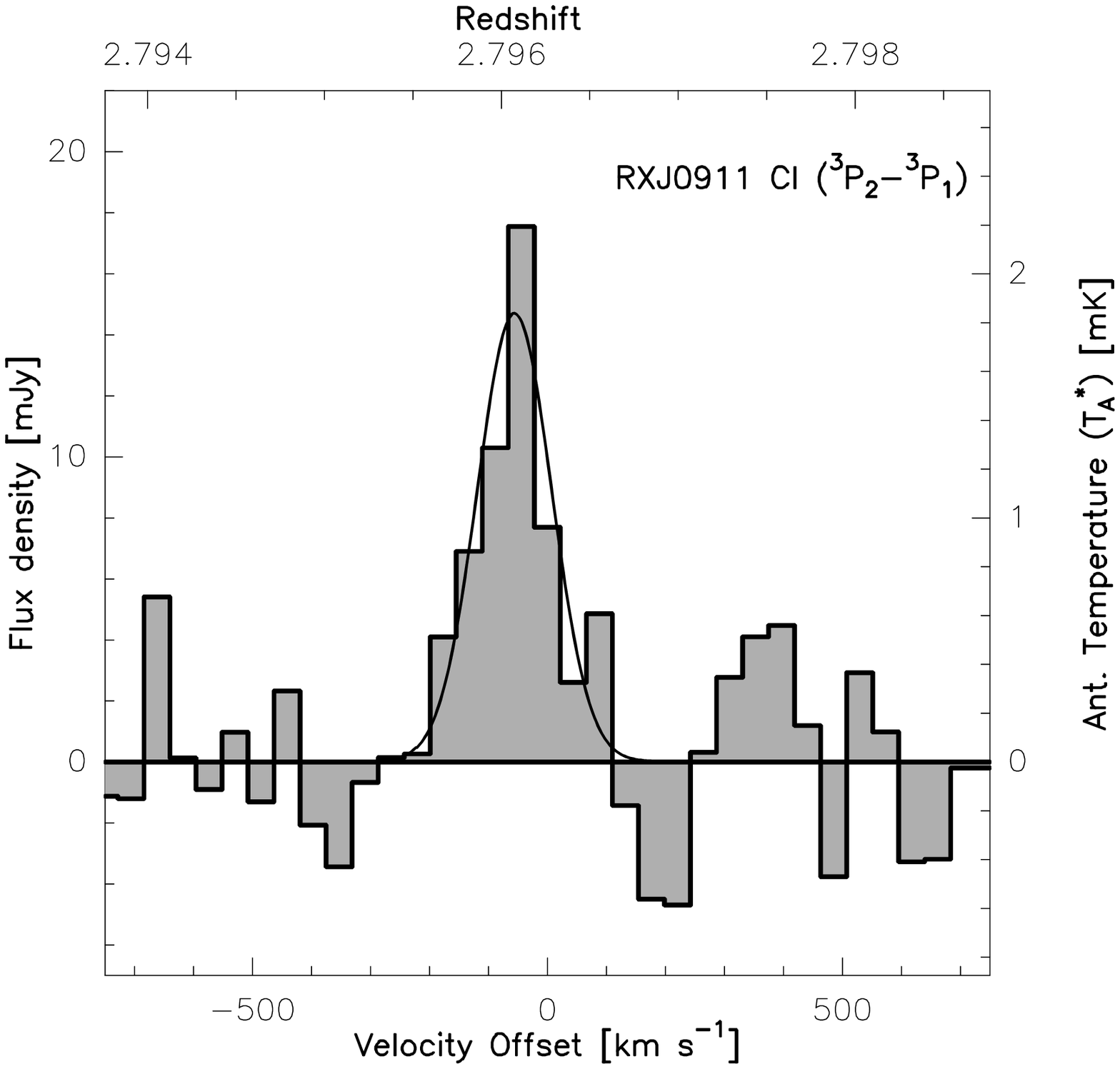}
\end{center}
\caption{{\em Left:} \cone\ spectrum of RX\,J0911+0551. {\em Right:} \ctwo\ spectrum of RX\,J0911+0551. See Table~3 for respective fluxes and noise.}
\end{figure*}

\begin{figure*}
\begin{center}
\includegraphics[width=7cm,angle=0]{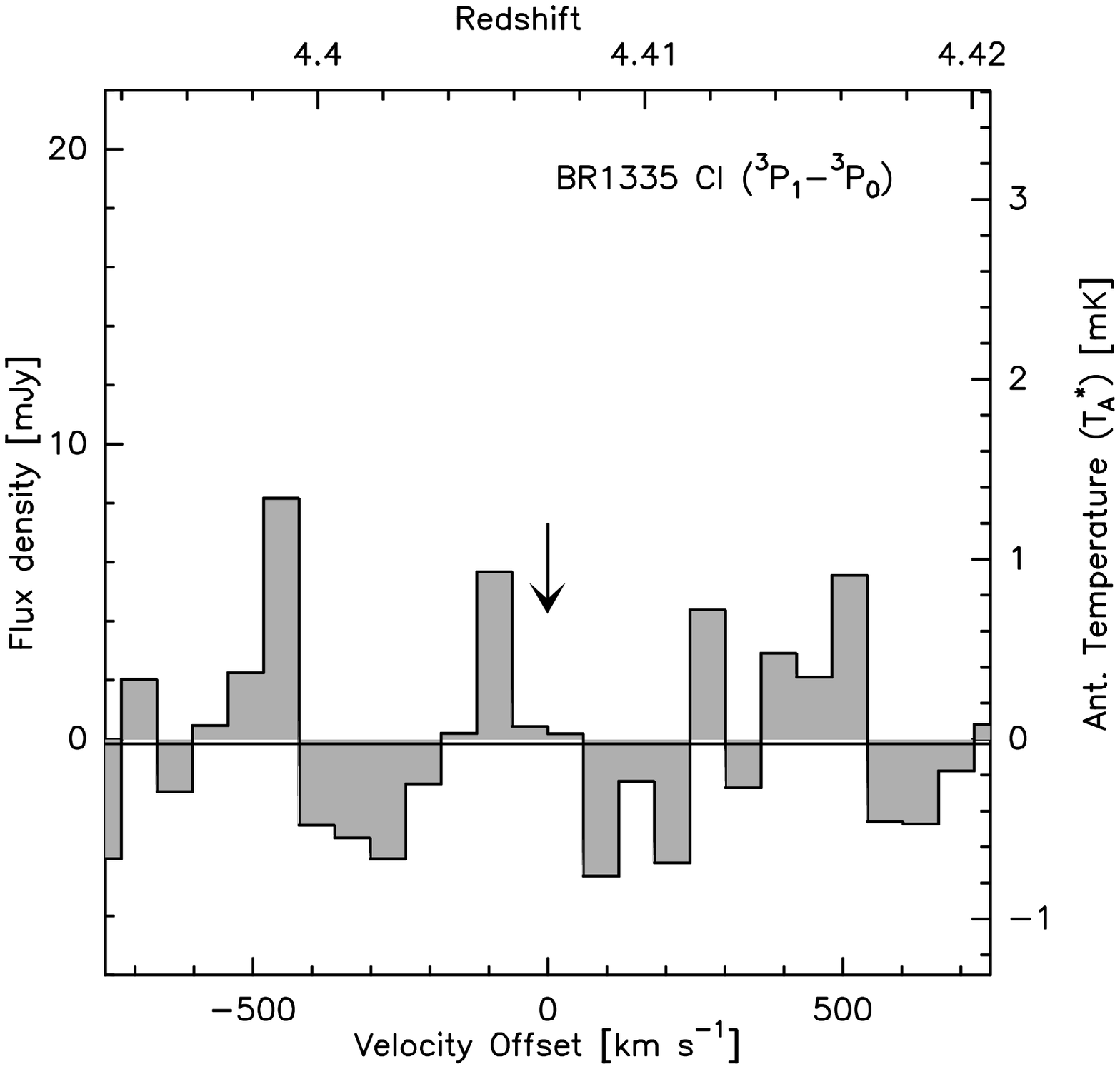}
\includegraphics[width=7cm,angle=0]{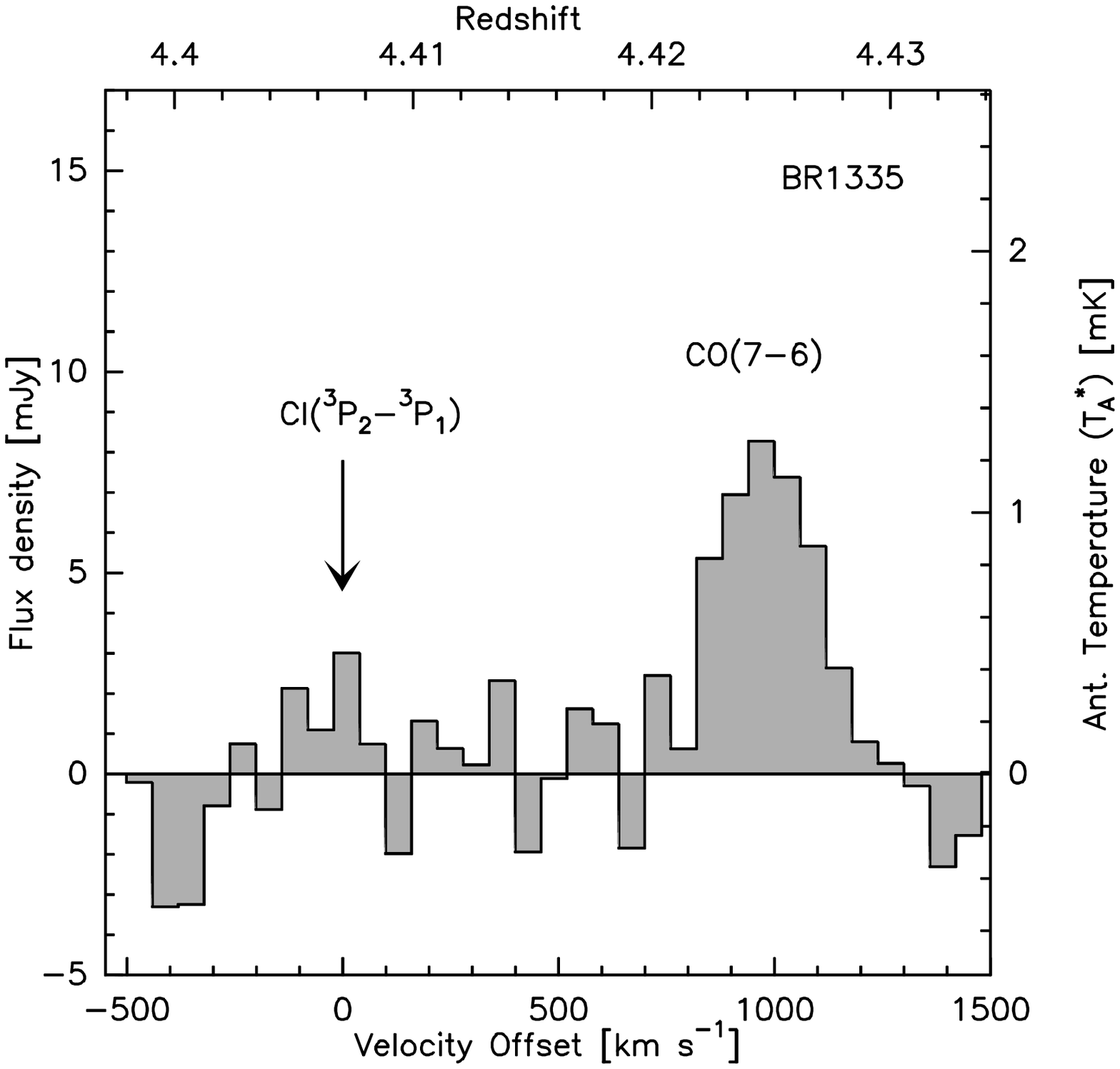}
\end{center}
\caption{\ci\ upper limits for BRI\,1335--0417. {\em Left:} \cone\ observations. {\em Right:} \ctwo\ observations including the adjacent CO(7--6) line. The arrows in both panels indicate the expected central position of the \ci\ lines.}
\end{figure*}

\subsection{30\,m observations}

We have observed both the \cone\ and \ctwo\ emission in 6 additional
sources, four quasars and two submillimeter galaxies. All fitted
parameters and line fluxes are summarized in Table~3. In the following
we briefly discuss the individual sources.

\subsubsection{APM\,08279+5255}

This source is one of the (apparently) brightest sources known to
date, with a submillimeter flux of S$_{850\mu m}$=75$\pm$4\,mJy (Lewis
et al.\ 1998). The first CO detections by Downes et al.\ (1999, $J$=4,
$J$=9) already indicated the extreme excitation conditions of the
molecular gas (studied in detail in Wei{\ss} et al.\ 2007). One of the
explanations for its apparent brightness and extreme excitation is
that the lensing geometry is such that the central 100\,pc of this
source is magnified by a large factor ($\mu$=60--100 Wei{\ss} et al.\
2007, see also Egami et al.\ 2000)\footnote{We note that Riechers et
al.\ 2009a derived a magnification factor of only $\sim$4 for the
molecular gas phase.}. The \cone\ line has been detected by Wagg et
al.\ (2006). The source is not detected in \ctwo\ emission with an
upper limit of 1\,Jy\,km\,s$^{-1}$ (Figure~7, left; upper limits are
given in Table~3).

\subsubsection{RX\,J0911+0551}

This source is bright in the submillimeter (S$_{850\mu
m}$=26.7$\pm$1.4\,mJy, Barvainis \& Ivison 2002) and CO has been
tentatively detected in this source by Hainline et al.\ (2004,
$J$=3). Follow--up observations by W11 ($J$=3,5,7,8,9) show that the CO
emission line is very narrow (FWHM$\sim$110\kms).  We detect both the
\cone\ and \ctwo\ emission lines in this source (see Figure~8 and
Table~3).

\subsubsection{BRI\,1335--0417}

BRI\,1335--0417 is bright in the submillimeter (e.g., Benford et al.\
1999, we adopt S$_{850\mu m}$=22$\pm$4\,mJy from their Figure~2) and
has first been detected in CO ($J$=5) by Guilloteau et al.\ (1997). High
angular resolution CO observations revealed a complex, possibly
merging system (Carilli et al.\ 2002a, Riechers et al.\ 2008a,
$J$=2). The source is not detected in both the \cone\ and \ctwo\ line
with upper limits of 2.2\,Jy\,km\,s$^{-1}$ and 0.8\,Jy\,km\,s$^{-1}$
(Figure~9, right; see also Table~3). The spectrum of the \ctwo\ line
in Figure 9 also includes the CO(7--6) line that is close in
frequency space (the CO emission is discussed in W11).

\begin{figure}
\begin{center}
\includegraphics[width=7cm,angle=0]{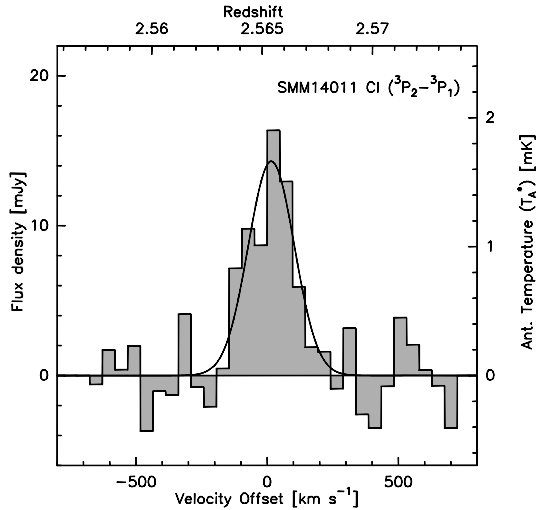}
\end{center}
\caption{ \ctwo\ spectrum of SMM\,J14011+0252. See Table~3 for
fluxes and noise (see Wei{\ss} et al., 2005b, for the \cone\ spectrum of the same source).}
\end{figure}

\begin{figure*}
\begin{center}
\includegraphics[width=7cm,angle=0]{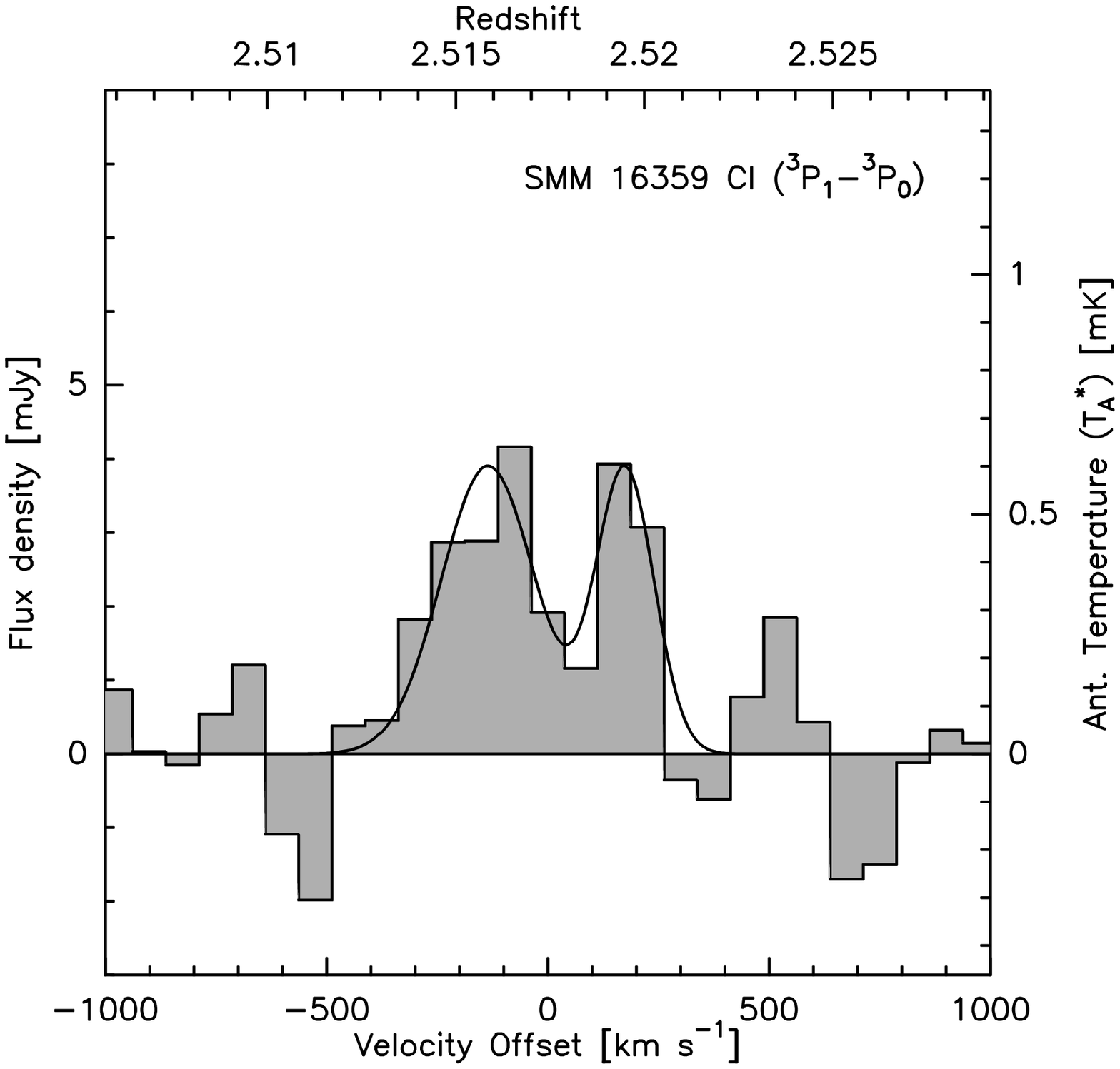}
\includegraphics[width=7cm,angle=0]{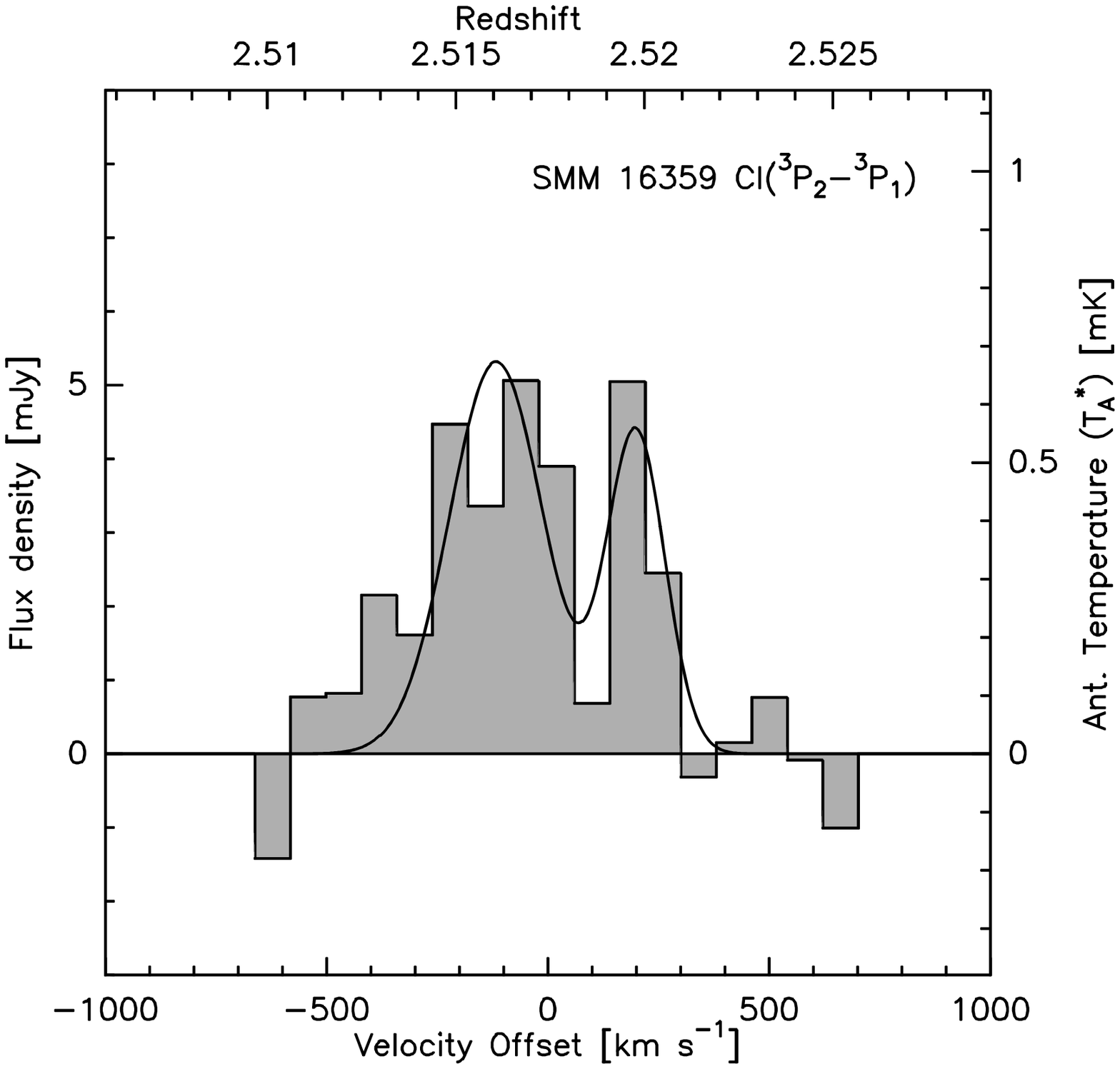}
\end{center}
\caption{{\em Left:} \cone\ spectrum of SMM\,J16359+6612, {\em Right:} \ctwo\ spectrum of SMM\,J16359+6612. Two Gaussian functions have been fitted to account for the two components (e.g., Sheth et al.\ 2004, Kneib et al.\ 2005, Wei{\ss} et al.\ 2005a). See Table~3 for fluxes and noise.}
\end{figure*}

\begin{figure}
\begin{center}
\includegraphics[width=7cm,angle=0]{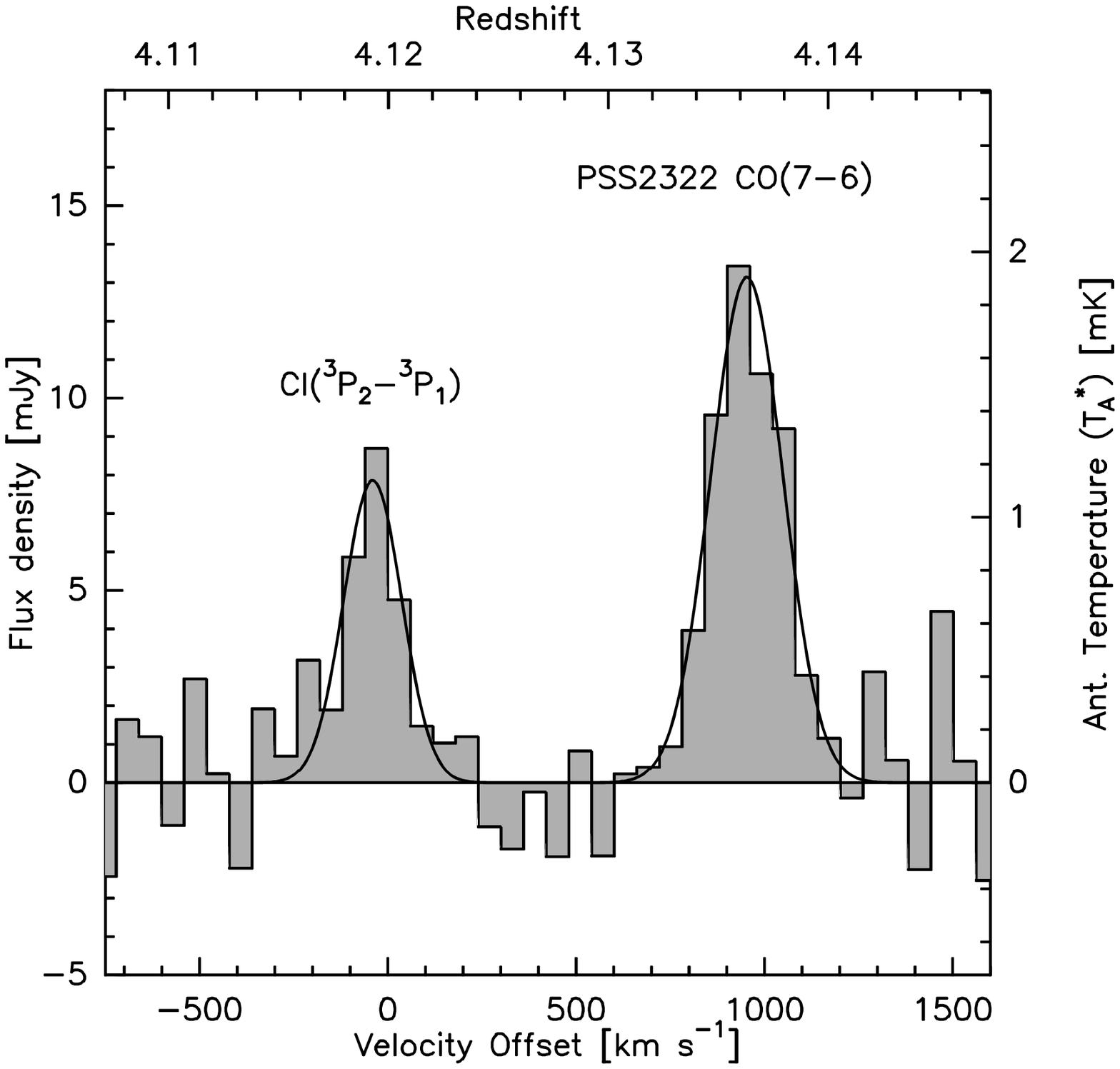}
\end{center}
\caption{\ctwo\ spectrum of PSS\,2322+1944 with the adjacent CO(7--6) line (W11). The velocity and redshift scales refer to the \ctwo\ observations. See Table~3 for fluxes and noise.}
\end{figure}

\subsubsection{SMM\,J14011+0252}

SMM\,J14011+0252 is one of the first submillimeter galaxies (S$_{850\mu
m}$=15$\pm$2\,mJy, Ivison et al.\ 2000) that has been detected in CO
emission (Frayer et al.\ 1999, $J$=3; Downes \& Solomon 2003, $J$=3,
$J$=7). Downes \& Solomon argue that the source is strongly lensed, with
a magnification factor $\mu$=25$\pm$5. The \ctwo\ line is shown in
Figure~10 (see also Table~3).

\subsubsection{SMM\,J16359+6612}

This source is a triple--lensed system in which all three images are
bright in the submillimeter (brightest image `B': S$_{850\mu
m}$=17$\pm$2\,mJy, Kneib et al.\ 2004). All images have subsequently
been detected in CO emission (Sheth et al.\ 2004, $J$=3, Kneib et al.\
2005, $J$=3). Kneib et al.\ (2004) have argued that the magnifications
for the three images are 14$\pm$2, 22$\pm$2 and 9$\pm$2,
respectively. Higher--$J$ CO transitions have been presented in Wei{\ss}
et al.\ (2005a). The beams are 18$''$ and 11$''$ at our tuned
frequencies at 2\,mm and 1\,mm respectively, i.e. only the brighest
image (`B') is covered in our observations. In Figure~11 we show the
\cone\ (left) and the \ctwo\ (right) spectra (see also Table~3).

\subsubsection{PSS\,J2322+1944}

PSS\,J2322+1944 is a lensed source ($\mu$=5.4$\pm$0.3, Riechers et
al.\ 2008b) which is bright in the submillimeter (S$_{850\mu
m}$=22.5$\pm$2.5\,mJy, Isaak et al.\ 2002; S$_{1300\mu
m}$=9.6$\pm$0.5\,mJy, Omont et al.\ 2001). Strong CO emission has been
reported by Cox et al.\ (2002, $J$=4, $J$=5) and resolved imaging has
revealed a molecular Einstein ring in this source by Carilli et al.\
(2002b, 2003, $J$=2) and Riechers et al.\ (2008b, $J$=2). The \cone\ line
has been detected by Pety et al.\ (2004). We detect the \ctwo\ line at
high significance (Table~3, Figure~12 shows a wider bandwidth spectrum
after combining the \ctwo\ data with the CO(7--6) data presented in
W11).

\begin{turnpage}
\begin{deluxetable*}{lllllll|llllll}
\tablecaption{Compilation of all high--z \ci\ observations to date}
\tablewidth{0pt}
\tablehead{
\colhead{source} & \colhead{z} & \colhead{magn. ($\mu$)} & \colhead{I$_{\rm CO(3-2)}$} & \colhead{I$_{\rm CI(1-0)}$} & \colhead{I$_{\rm CI(2-1)}$} & \colhead{S$_{\rm 850\mu m}$} & \multicolumn{6}{l}{references for columns\tablenotemark{a}}\\
\colhead{} & \colhead{} & \colhead{} & \colhead{[Jy\,km\,s$^{-1}$]} & \colhead{[Jy\,km\,s$^{-1}$]} & \colhead{[Jy\,km\,s$^{-1}$]} & [mJy]  &  \colhead{[2]} &  \colhead{[3]} &  \colhead{[4]} &  \colhead{[5]} &  \colhead{[6]} &  \colhead{[7]}\\
\colhead{[1]} & \colhead{[2]} & \colhead{[3]} & \colhead{[4]} & \colhead{[5]} & \colhead{[6]} & \colhead{[7]} &  \multicolumn{6}{l}{}
}
\startdata
SMM\,J02399--0136& 2.808               &  2.5                      & 3.1$\pm$0.4                  &  1.9$\pm$0.2  &  1.8$\pm$0.9                  &        26$\pm$3                  & 1,2      & 1,2   & 2        & 0     &  0 & 3     \\
APM\,08279+5255  & 3.911                  & 60--100\tablenotemark{b}  & 2.2$\pm$0.2\tablenotemark{c} & 0.93$\pm$0.13 & $<$1.1                        &        75$\pm$4\tablenotemark{d} & 4        & 5     & 4        & 6     &  0 & 7,8   \\
RX\,J0911+0551   & 2.796                  & $\sim$20                  & 2.0$\pm$0.3\tablenotemark{e} &  2.1$\pm$0.3  &  2.3$\pm$0.4                  &      26.7$\pm$1.4                & 9,10     & 8     & 9,10     & 0     &  0 & 8     \\         
F10214           & 2.285                  & 10\tablenotemark{f}       & 4.2$\pm$0.8                  &  2.0$\pm$0.4  &  4.6$\pm$0.7\tablenotemark{g} &        50$\pm$5                  & 11       & 11    & 11       & 12,13 & 13 & 8     \\
SDSS\,J1148+5251 & 6.419                  &  1                        & 0.18$\pm$0.04                &   ---         & 0.22$\pm$0.05                 &       7.8$\pm$0.7                & 14,15    & ---   & 15,16    & ---   & 17 & 18    \\
SMM\,J123549+6215& 2.202                  &  1                        & 1.6$\pm$0.2                  &  1.1$\pm$0.2  & $<$1.0                        &       8.3$\pm$2.5                & 19,20    & ---   & 19,20    & 0     &  0 & 21    \\
GN\,20.2         & 4.051                  &  1                        & 0.5$\pm$0.2\tablenotemark{c} &  ---          & $<$1.9\tablenotemark{h}       & $\sim$9.9$\pm$2.3                & 22       & 23    & 22       & ---   & 24 & 25    \\
GN\,20           & 4.055                  &  1                        & 0.8$\pm$0.1\tablenotemark{c} &  ---          & $<$1.2\tablenotemark{h}       &      20.3$\pm$2.1                & 22       & 23    & 22       & ---   & 24 & 25    \\
BRI\,1335--0417  & 4.407                  &  1                        & 1.0$\pm$0.2\tablenotemark{i} &  $<2.2$       & $<$0.8                        &        23$\pm$3\tablenotemark{j} & 26       & ---   & 9        & ---   &  0 & 27    \\
SMM\,J14011+0252 & 2.565                  & 25$\pm$5\tablenotemark{k} & 2.8$\pm$0.3                  &  1.8$\pm$0.3  &  3.1$\pm$0.3                  &        15$\pm$2                  & 28,29    & 29    & 28,29    & 12    &  0 & 30    \\
Cloverleaf       & 2.558                  & 11\tablenotemark{l}       & 13.2$\pm$0.2                 &  3.9$\pm$0.6  &  5.2$\pm$0.3                  &      58.8$\pm$8.1                & 12       & 31    & 32       & 12,33 & 32 & 8     \\
SMM\,J16359+6612 & 2.517\tablenotemark{m} & 22$\pm$2                  & 2.8$\pm$0.2                  &  1.7$\pm$0.3  &  1.6$\pm$0.3                  &        17$\pm$2                  & 12,34,35 & 35    & 12,34,35 & 0     &  0 & 36    \\
SMM\,J163650+4057 & 2.385                  &  1                        & 1.6$\pm$0.2                  &  $<$0.72      & $<$2.0                        &       8.2$\pm$1.7                & 37,19,20 & ---   & 37,19,20 & 0     &  0 & 38    \\
SMM\,J163658+4105 & 2.452                  &  1                        & 1.8$\pm$0.2                  & 0.90$\pm$0.2  & $<$2.0                        &      10.7$\pm$2.0                & 39,19,20 & ---   & 37,19,20 & 0     &  0 & 38    \\
MM\,18423+5938    & 3.930                  &  20\tablenotemark{n}  &  $2.8\pm0.3$c                 &  $2.3\pm0.5$ &  $4.2\pm0.8$                   &   $70\pm$5\tablenotemark{o}      & 40       & ---   & 40       & 40    & 40 & 40   \\
SMM\,J213511-0102 & 2.326                  & $32.5\pm4.5$             & $13.2\pm0.1$                 & $16.0\pm0.5$  & $16.2\pm0.6$                  &  $113\pm13$                       & 41      & 42     & 43       & 43    & 43 & 42    \\
PSS\,J2322+1944  & 4.120                  & 5.3$\pm$0.3               & 2.1$\pm$0.3\tablenotemark{c} & 0.80$\pm$0.12 &  1.4$\pm$0.3                  &      22.5$\pm$2.5                & 44       & 45    & 44       & 46    &  0 & 47    
\enddata
\tablenotetext{a}{References in last column are the following papers:
[0] this work; 
[1] Frayer et al.\ 1998; 
[2] Genzel et al.\ 2003; 
[3] Ivison et al.\ 1998; 
[4] Downes et al.\ 1999; 
[5] Wei{\ss} et al.\ 2007;
[6] Wagg et al.\ 2006; 
[7] Lewis et al.\ 1998; 
[8] Barvainis \& Ivison 2002; 
[9] W11;
[10] Hainline et al.\ 2004; 
[11] Downes et al.\ 1995; 
[12] Wei{\ss} et al.\ 2005b; 
[13] Ao et al.\ 2008; 
[14] Bertoldi et al.\ 2003; 
[15] Walter et al.\ 2003; 
[16] Walter et al.\ 2004; 
[17] Riechers et al.\ 2009b; 
[18] Robson et al.\ 2004; 
[19] Tacconi et al.\ 2006; 
[20] Tacconi et al.\ 2008; 
[21] Chapman et al.\ 2003; 
[22] Daddi et al.\ 2009a;
[23] Daddi et al.\ 2009b;
[24] Casey et al.\ 2009; 
[25] Pope et al.\ 2006;
[26] Guilloteau et al.\ 1997;
[27] Benford et al.\ 1999;
[28] Frayer et al.\ 1999; 
[29] Downes \& Solomon 2003; 
[30] Ivison et al.\ 2000; 
[31] Venturini \& Solomon 2003;
[32] Wei{\ss} et al.\ 2003; 
[33] Barvainis et al.\ 1997; 
[34] Sheth et al.\ 2004; 
[35] Kneib et al.\ 2005; 
[36] Kneib et al.\ 2004; 
[37] Neri et al.\ 2003; 
[38] Ivison et al.\ 2002; 
[39] Greve et al.\ 2005; 
[40] Lestrade et al.\ 2010;
[41] Swinbank et al.\ 2010;
[42] Ivison et al.\ 2010c;
[43] Danielson et al.\ 2010;
[44] Cox et al.\ 2002; 
[45] Riechers et al.\ 2008b;
[46] Pety et al.\ 2004; 
[47] Isaak et al.\ 2002. 
}
\tablenotetext{b}{An average value of 80 is used for further analysis. See Riechers et al.\ (2009a) for a significantly lower $\mu$ for the CO emission.}
\tablenotetext{c}{No CO(3--2) measurements available. CO(3--2) fluxes have been derived from measured CO(4--3) fluxes assuming that the emission is thermalized up to $J$=4 (i.e., S$_{\rm CO(3-2)}$=$(\frac{3}{4})^2\times$S$_{\rm CO(4-3)}$.}
\tablenotetext{d}{Barvainis \& Ivison give S$_{\rm 850\mu m}$=84$\pm$3\,mJy.}
\tablenotetext{e}{Data from W11. Hainline et al.\ (2004) reported a significantly larger velocity width/total flux.}
\tablenotetext{f}{Downes \etal\ (1995) derive slightly different magnifications for the CO ($\mu$=10) and FIR ($\mu$=13) emission.}
\tablenotetext{g}{Consistent with the 2$\sigma$ upper limit of 7\,Jy\,km\,s$^{-1}$ given by Papadopoulos (2005).}
\tablenotetext{h}{2$\sigma$ upper limits by Casey et al.\ (2009).}
\tablenotetext{i}{Extrapolated from the $J$=5 measurement by Guilloteau et al.\ (1997).}
\tablenotetext{j}{Value taken from Figure~2 of Benford et al.\ (1999).}
\tablenotetext{k}{Ivison \etal\  (2001) adopted a magnification value that is lower by an order of magnitude ($\mu$=2.5).}
\tablenotetext{l}{Kneib \etal\ (1998) give a magnification range of $\mu$=18--30.}
\tablenotetext{m}{Only the brightest component (`B') is covered here.}
\tablenotetext{n}{Magnification factor unknown (Lestrade et al.\ 2010). We here assign a factor of 20 to this source; consistent with the large factors found for other strongly lensed SMGs.}
\tablenotetext{o}{The 850 micron flux is derived from the 1.2\,mm flux given in Lestrade et al.\ 2010, assuming a modified
black body with T$_{\rm dust}$=40\,K and $\beta$=1.6.}
\end{deluxetable*}
\end{turnpage}

\section{Discussion}

\subsection{Previous high--z \ci\ observations}

We have observed \ci\ emission in 10 high--redshift systems, targeting
a total of 16 \ci\ lines, and have detected a total of 10 lines (one
more tentatively). In the discussion that follows we include \ci\
measurements of high--redshift sources from the literature (an
additional 7 sources). All \ci\ measurements (including upper limits)
to date are summarized in Table~4. We have re--calculated all line and
far--infrared luminosities for the published sources to ensure that
consistent assumptions and cosmological parameters are used in the
analysis that follows.

\newpage
\subsection{\ci, CO and FIR luminosities}

\subsubsection{Line luminosities}

In the following we derive line luminosities both in units of
$L_\odot$ ($L_{\rm line}$) and K\,km\,s$^{-1}$\,pc$^2$ ($L'_{\rm
line}$), where `line' refers to either CO(3--2), \cone\ or \ctwo. The
former luminosity is given by

\begin{equation}
L_{\rm line} = 1.04 \times 10^{-3} \, S_{\rm line}\, \Delta v\,
     \nu_{\rm rest} (1+z)^{-1}\, D_{\rm L}^{2},
\label{eq1}
\end{equation}

(Solomon et al.\ 1992), where the luminosity of the line, $L_{\rm line}$, is
measured in $L_\odot$; the velocity integrated flux, $S_{\rm line}\,
\Delta v$, in Jy\,\kms; the rest frequency, $\nu_{\rm rest} = \nu_{\rm
obs} (1+z)$, in GHz; and the luminosity distance, $D_{\rm L}$, in Mpc
(we have tabulated the luminosity distances in Table~5 assuming the
cosmological parameters given at the end of Section~1). $L_{\rm line}$
(Eq.~1) is a true luminosity, i.e. it presents the amount of cooling
that is radiated away by the respective line.

The line luminosity can also be expressed in units of
K\,km\,s$^{-1}$\,pc$^{2}$:

\begin{equation}
L'_{\rm line} = 3.25 \times 10^{7} \, S_{\rm line}\, \Delta v \,
     \nu_{\rm obs}^{-2}\, D_{\rm L}^{2} \, (1+z)^{-3}.
\label{eq3}
\end{equation}

(Solomon et al.\ 1992).  As noted by Solomon \& Vanden Bout (2005), as
L$'_{\rm line}$ is proportional to the brightness temperature, the
L$'$ ratio of two lines is a measure of the ratio of their intrinsic
brightness temperatures (assuming that both lines arise from the same
volume, i.e. same area and linewidth).

We have calculated both luminosities for the \cone, \ctwo\ and
CO(3--2) lines and they are given in Table~5 (uncorrected for possible
magnification).

\subsubsection{Far--Infrared (FIR) luminosities}

To derive consistent results within our sample, we have recalculated
FIR luminosites for all sources based on their measured 850$\mu$m flux
(as tabulated in Table~4). The reason why we chose the 850$\mu$m band
as our reference wavelength is because continuum emission has been
measured in this band for almost every source in our sample.

Following Blain et al.\ (2003), a simple description of the
far--infrared SED as a function of frequency $\nu$ is based on a
backbody spectrum $B_\nu\sim\nu^3/[exp(h\nu/kT)-1]$ at temperature
$T$, modified by a frequency--dependent emissivity function
$\epsilon_\nu\sim\nu^\beta$ (`modified black body').  This yields a
functional behavior at the source's rest frame of $f_\nu \sim
\nu^{3+\beta}/[exp(h\nu/kT)-1]$ (e.g. Blain et al.\ 2003).

At wavelengths where the dust is optically thin, the observed flux is
proportional to the dust mass and the the observed flux density can be
written as:

$$
S_{\nu_\mathrm{o}} = \frac{(1+z)}{{D_\mathrm{L}}^2}\ M_\mathrm{dust} \kappa(\nu_\mathrm{r}) B_{\nu_\mathrm{r}}(T_\mathrm{dust})\\ 
$$

(Downes et al.\ 1992), where $\kappa(\nu)$ is the dust absorption
coefficient and the subscripts $o$ and $r$ refer to the observed and
restframe frequencies.

We have derived FIR luminosities by integrating under a modified
blackbody curve from 40 to 400\,$\mu$m (Sanders et al.\
2003)\footnote{The FIR luminosities decrease by 20--30\% if the
definition by Helou et al.\ 1985 [42--122$\mu$m] is used.}. The
numbers are given in the last column in Table~5. For the quasars in
our sample we have used the best--fit values (T$_{\rm
dust}$=47$\pm$3\,K, $\beta$=1.6$\pm$0.1) derived by Beelen et al.\
(2006). Likewise, we have used the best--fit values by Kov{\'a}cs et
al.\ (2006) for the SMGs (T$_{\rm dust}$=34.6$\pm$3\,K,
$\beta$=1.5). In both cases, following Beelen et al. (2006) and
Kov{\'a}cs et al.\ (2006) we assign uncertainties of $\pm3$\,K to
T$_{\rm dust}$ and $\pm$0.1 to $\beta$. Based on this we assign
uncertainties of $\Delta$(log $L_{\rm FIR}$)=0.15, corresponding to
$\sim40\%$, to our FIR luminosity estimates (the measurement
uncertainty in the 850$\mu$m fluxes is significantly lower than this
number).

All luminosities given in Table~5 have {\em not} been corrected for
magnification. To derive the unlensed luminosities, all numbers need
to be divided by the respective magnification factors ($\mu$) given
in Table~4.

\begin{deluxetable*}{lllllllll}
\tablecaption{Derived Luminosities based on the values given in Table~4.}
\tablewidth{0pt}
\tablehead{
\colhead{source} & 
\colhead{D$_L$\tablenotemark{a}} & 
\colhead{L$_\cone$\tablenotemark{b}} & 
\colhead{L$_\ctwo$\tablenotemark{b}} & 
\colhead{L$_{\rm CO(3-2)}$\tablenotemark{b}} & 
\colhead{L$'_\cone$\tablenotemark{c}} & 
\colhead{L$'_\ctwo$\tablenotemark{c}} & 
\colhead{L$'_{\rm CO(3-2)}$\tablenotemark{c}}  & 
\colhead{L$_{\rm FIR}$\tablenotemark{d}}\\
\colhead{} & 
\colhead{[Gpc]} & 
\colhead{[10$^7$\,L$_\odot$]} & 
\colhead{[10$^7$\,L$_\odot$]} & 
\colhead{[10$^7$\,L$_\odot$]}  & 
\colhead{[10$^{10}$\,K\,km\,s$^{-1}$\,pc$^{2}$]}  & 
\colhead{[10$^{10}$\,K\,km\,s$^{-1}$\,pc$^{2}$]} & 
\colhead{[10$^{10}$\,K\,km\,s$^{-1}$\,pc$^{2}$]} & 
\colhead{10$^{13}$\,L$_\odot$}
}
\startdata
  SMMJ02399-0136 & 23.84 &    14.5$\pm$1.5 &    22$\pm$11 &    17$\pm$2    &	3.8$\pm$0.4 &	 1.3$\pm$0.7  &  12.6$\pm$1.6 &  1.41\\
   APM08279+5255 & 35.55 &    12.3$\pm$1.7 & $<$24        &  20.4$\pm$1.9  &	3.2$\pm$0.5 & $<$1.4	      &  15.4$\pm$1.4 &  9.45\\
    RXJ0911+0551 & 23.72 &      16$\pm$2   &    29$\pm$4  &  10.7$\pm$1.4  &	4.2$\pm$0.6 &	 1.7$\pm$0.3  &   8.1$\pm$1.1 &  3.81\\
          F10214 & 18.53 &    10.9$\pm$1.9 &    40$\pm$6  &    16$\pm$3    &	2.9$\pm$0.5 &	 2.4$\pm$0.4  &    12$\pm$2   &  7.58\\
  SDSSJ1148+5251 & 63.79 &    \nodata	   &    10$\pm$2  &   3.6$\pm$0.8  &     \nodata    &	0.60$\pm$0.14 &   2.7$\pm$0.6 &  0.87\\
 SMMJ123549+6215 & 17.71 &     5.5$\pm$1.1 & $<$8.2       &   5.6$\pm$0.7  &	1.5$\pm$0.3 & $<$0.5	      &   4.3$\pm$0.5 &  0.45\\
          GN20.2 & 37.08 &    \nodata      & $<$44        &     5$\pm$2    &	 \nodata    & $<$2.6	      &   3.7$\pm$1.5 &  0.55\\
            GN20 & 37.12 &    \nodata      & $<$28        &   7.8$\pm$1.0  &	 \nodata    & $<$1.6	      &   5.9$\pm$0.7 &  1.12\\
    BRI1335-0417 & 40.99 &  $<$35          & $<$21        &    11$\pm$2    & $<$9.2         & $<$1.2	      &   8.5$\pm$1.7 &  2.78\\
  SMMJ14011+0252 & 21.35 &     12$\pm$2    &    33$\pm$3  &  12.9$\pm$1.4  &	3.1$\pm$0.5 &	1.97$\pm$0.19 &   9.7$\pm$1.0 &  0.82\\
      Cloverleaf & 21.28 &     25$\pm$4    &    56$\pm$3  &  60.4$\pm$0.9  &	6.7$\pm$1.0 &	3.28$\pm$0.19 &  45.7$\pm$0.7 &  8.64\\ 
  SMMJ16359+6612 & 20.86 &   10.8$\pm$1.9  &    23$\pm$3  &  12.5$\pm$0.9  &	2.8$\pm$0.5 &	1.35$\pm$0.18 &   9.4$\pm$0.7 &  0.93\\
  SMMJ163650+4057 & 19.53 &  $<$4.2        & $<$19        &   6.5$\pm$0.8  & $<$1.1	    & $<$1.1	      &   4.9$\pm$0.6 &  0.45\\
  SMMJ163658+4105 & 20.21 &    5.5$\pm$1.1 & $<$20        &   7.7$\pm$0.9  &	1.4$\pm$0.3 & $<$1.2	      &   5.8$\pm$0.6 &  0.58\\
  MM18423+5938    & 35.76 &   $30\pm7$     & $92\pm17$    & $26\pm3$       & $8.0\pm1.7$    & $5.4\pm1.0$     & $20\pm2$      &   3.84 \\
SMMJ213511-0102   & 18.94 &   $88\pm3$     & $147\pm5$    & $51.2\pm0.4$   & $23.2\pm0.7$   & $8.7\pm0.3$     & $38.7\pm0.3$  &   6.15 \\
   PSSJ2322+1944  & 37.83 &   11.5$\pm$1.7 &    33$\pm$7  &    21$\pm$3    &	3.0$\pm$0.5 &	 1.9$\pm$0.4  &    16$\pm$2   &   2.78
\enddata
\tablecomments{All luminosities given in this table are {\em uncorrected} for lensing. To correct for gravitational magnification, all numbers have to be divided by $\mu$ given in Table 4, Column 3.}
\tablenotetext{a}{For $H_{\rm 0} = 71$ \kms\,Mpc$^{-1}$,$\Omega_\Lambda=0.73$ and $\Omega_m=0.27$ (Spergel \etal\ 2003, 2007).}
\tablenotetext{b}{From Equation~1, all units in 10$^7$\,L$_\odot$.}
\tablenotetext{c}{From Equation~2, all units in 10$^{10}$\,K\,km\,s$^{-1}$\,pc$^{2}$.}
\tablenotetext{d}{See discussion in Section~4.2.2.}
\end{deluxetable*}

\subsection{\ci, CO and \ci/CO line ratios}

In Figure~13 we plot the \cone\ and \ctwo\ line luminosities as a
function of CO(3--2) luminosities. Most \ci\ luminosities are between
10\% and 100\% of the CO(3--2) luminosities (as indicated by the
dashed diagonal lines). We calculate the $L'_\cone$/$L'_{\rm CO(3-2)}$
ratios of all targets (Table~6). We chose the CO(3--2) transition as a
measure for the CO luminosity as this transition has been observed for
almost all targets.  We find an average value of $L'_\cone$/$L'_{\rm
CO(3-2)}$=0.32$\pm$0.13 with no significant differences between the
QSO and SMG sample. We plot the $L'_\cone$/$L'_{\rm CO(3-2)}$ ratios
in Figure~14 as a function of $L_\cone$/$L_{\rm FIR}$ (discussed in
Section~4.7).

We now compare the $L'_\cone$/$L'_{\rm CO(3-2)}$ ratios to systems at
low--redshift (including our Galaxy). This comparison is complicated
by a two facts: (a) in most low--z cases, the CO luminosity has been
derived from the $J$=1 transition, not $J$=3 as in this study. We here
assume $L'_{\rm CO(1-0)}\!\approx0.9\times L'_{\rm CO(3-2)}$, Wei{\ss}
et al.\ 2005a, Ao et al.\ 2008) -- this translates to a
$L'_\cone$/$L'_{\rm CO(1-0)}$=0.29$\pm$0.12 for the high--redshift
sample; (b) The beam sizes of the \cone\ and CO observations are
typically different for the low--z sources (this fact does not play a
role at high redshift, where our beam sizes are in most cases larger
than the source size).

Gerin \& Phillips (2000) found an average value of $L'_\cone$/$L'_{\rm
CO(1-0)}$=$0.2\pm0.2$ in nearby galaxies and similar values have been
derived in other studies: e.g., M\,83: 0.18$\pm$0.04, NGC\,6946:
0.20$\pm$0.04 (Israel \& Baas 2001, their table~4), IC\,342:
0.16$\pm$0.03, Maffei~2: 0.10$\pm$0.05 (Israel \& Baas 2003, their
table~4), M\,51: 0.22$\pm$0.06 (Israel et al.\ 2006, their table~4),
Hennize\,2--10: 0.09$\pm$0.03, NGC\,253: 0.22$\pm$0.05 (Bayet et al.\
2004, their tables 2 and 3), M\,82: 0.2\footnote{Using the \cone\
measurement from Stutzki et al.\ 1997 and the CO(1--0) data presented
in Walter et al.\ 2002.}. The average of the same value in the Milky
Way is around 0.15$\pm$0.1; Fixsen et al.\ (1999) (0.09, Ojha et al.\
(2001): 0.08, Oka et al.\ (2005): ranging from 0.05--0.3).

We conclude that, to first order, the $L'_\cone$/$L'_{\rm CO}$ ratio
does not appear to vary strongly with either environment or redshift,
from low--excitation environments like the Milky Way all the way to
extreme starbursts in the early universe.  This finding is in
agreement with our earlier conclusion based on a significantly smaller
sample (Wei{\ss} et al.\ 2005b).

\begin{deluxetable*}{llllllll}
\tablecaption{Derived Physical Properties.}
\tablewidth{0pt}
\tablehead{
\colhead{source} & 
\colhead{T$_{\rm ex}$\tablenotemark{a}} & 
\colhead{M$_\ci$} & 
\colhead{M$_{\rm H2}$} & 
\colhead{X[\ci]/X[H$_2$]} & 
\colhead{L$_\cone$/L$_{\rm FIR}$} & 
\colhead{L$'_\cone$/L$'_{\rm CO(3-2)}$} & 
\colhead{L$'_\ctwo$/L$'_\cone$}\\
\colhead{} & 
\colhead{[K]} & 
\colhead{[10$^7$\,M$_\odot$]} & 
\colhead{[10$^{10}$\,M$_\odot$]}  & 
\colhead{$10^{-5}$} & 
\colhead{10$^{-6}$} & 
\colhead{} & 
\colhead{}
}
\startdata
  SMMJ02399-0136 & 21.6$\pm$6.3  &  2.0$\pm$0.2  &   4.0$\pm$0.5   &   8.1$\pm$1.4 &     10$\pm$4   &   0.30$\pm$0.05 & 0.35$\pm$0.18 \\
   APM08279+5255 & 30.00	 & 0.050$\pm$0.007 & 0.154$\pm$0.014 &   5.4$\pm$0.9 &    1.3$\pm$0.6 &   0.21$\pm$0.03 & $<$0.44	  \\
    RXJ0911+0551 & 23.5$\pm$3.0  &  0.26$\pm$0.04  &  0.32$\pm$0.04  &    14$\pm$3   &    4.2$\pm$1.8 &   0.52$\pm$0.10 & 0.41$\pm$0.08 \\
          F10214 & 42.0$\pm$10.9   &  0.36$\pm$0.06  &  0.95$\pm$0.18  &   6.4$\pm$1.7 &    1.4$\pm$0.6 &   0.24$\pm$0.06 &  0.8$\pm$0.2  \\
  SDSSJ1148+5251 & 30.00	   & \nodata	     &   2.2$\pm$0.5   &  \nodata      &   \nodata      &  \nodata	  & \nodata	  \\
 SMMJ123549+6215 & 30.00	   &   1.8$\pm$0.3   &   3.4$\pm$0.4   &     9$\pm$2   &     12$\pm$5   &   0.34$\pm$0.08 & $<$0.34	  \\
          GN20.2 & 30.00	   & \nodata	     &   3.0$\pm$1.2   &  \nodata      & \nodata        &   \nodata	  &  \nodata	  \\
            GN20 & 30.00	   & \nodata	     &   4.7$\pm$0.6   &  \nodata      & \nodata        &   \nodata	  &   \nodata     \\
    BRI1335-0417 & 30.00	   & $<$11	     &   6.8$\pm$1.4   & $<$28         &  $<$13   & $<$1.1          & $<$0.13	  \\  
  SMMJ14011+0252 & 32.4$\pm$5.2   &  0.15$\pm$0.03  &  0.31$\pm$0.03  &   8.2$\pm$1.6 &     14$\pm$6   &   0.32$\pm$0.06 & 0.64$\pm$0.12 \\
      Cloverleaf & 26.7$\pm$3.0  &  0.76$\pm$0.12  &  3.32$\pm$0.05  &   3.8$\pm$0.6 &    2.9$\pm$1.3 &   0.15$\pm$0.02 & 0.49$\pm$0.08 \\
  SMMJ16359+6612 & 26.2$\pm$4.0  &  0.16$\pm$0.03  &  0.34$\pm$0.02  &   7.8$\pm$1.5 &     12$\pm$5   &   0.30$\pm$0.06 & 0.48$\pm$0.11 \\
  SMMJ163650+4057 & 30.00	   & $<$1.4          &   3.9$\pm$0.5   &   $<$5.8     &     $<$9       & $<$0.22	  & $<$1.0        \\
  SMMJ163658+4105 & 30.00	   &   1.8$\pm$0.4   &   4.6$\pm$0.5   &   6.4$\pm$1.5 &      9$\pm$4   &   0.25$\pm$0.06 & $<$0.82	  \\
  MM18423+5938    & $34.1\pm8.8$ &    $0.5\pm0.1$    &  $0.79\pm0.08$ &  $11\pm3$   & $8\pm4$ & $0.41\pm0.10$ & $0.7\pm0.2$ \\
 SMMJ213511-0102 & $22.4\pm0.6$   &  $0.91\pm0.03$  & $0.95\pm0.01$ & $15.9\pm0.5$ & $14\pm6$ & $0.60\pm0.02$ & $0.37\pm0.02$ \\
   PSSJ2322+1944 & 32.8$\pm$7.4   &  0.71$\pm$0.11  &   2.4$\pm$0.3   &   4.9$\pm$1.0 &    4.1$\pm$1.8 &   0.19$\pm$0.04 & 0.65$\pm$0.17 
\enddata
\tablecomments{All masses given in this table are {\em corrected} for lensing using $\mu$ given in Table 4, Column 3.}
\tablenotetext{a}{See Section~4.4. A temperature of 30\,K has been assumed for all objects for which T$_{\rm ex}$ could not be derived.}
\end{deluxetable*}

\subsection{Atomic Carbon Excitation Temperature}

Some sources have now been detected in both \ci\ lines and we can use
their flux ratio to derive the carbon excitation temperature.  As
discussed in Schneider et al.\ (2003) the carbon excitation
temperature in the local thermodynamic equilibrium (LTE) can be
derived from the \ci\ line ratio via the formula 

\begin{equation}
T_{\rm ex}\,=\,38.8\times ln(\frac{2.11}{\rm R})^{-1}, 
\end{equation}

where $R$\,=$L'_\ctwo$/$L'_\cone$. This assumes that both carbon lines share
the same excitation temperature and are optically thin. The latter
assumption can be tested using the equations given in Schneider et
al.\ (2003, their equations A.6 and A.7). Since these equations
require knowledge of T$_{\rm ex}$ we assume here, as a first guess,
T$_{\rm ex}$=T$_{\rm dust}$ (Sec.~4.2.2) and intrinsic source sizes of
r=1\,kpc. This yields optical depths for the \cone\ line between
0.05--0.3, justifying the use of equation~3. Larger source sizes, as
discussed by Tacconi et al.\ (2008), will lead to even lower optical
depths.

We summarize the derived excitation temperatures in Table~6 and find
values between $\sim$25 and $\sim$45\,K with an average value of
29.1$\pm$6.3\,K.  Note that if we used these derived T$_{\rm ex}$, the
\ci\ lines are still optically thin. In the local universe, the carbon
excitation tempertature could only be derived for a few objects to
date (using the same method as employed here): M\,82: $\sim$55\,K
(using the values given in Stutzki et al.\ 1997), NGC\,253:
$\sim$20\,K (Bayet et al.\ 2004), Milky Way: 17.5\,K (Fixsen et al.\
1999). Herschel SPIRE FTS observations will soon provide measurements
of both \ci\ lines in a sample of nearby ULIRGS (Van der Werf et al.\
2010).

\begin{figure*}
\begin{center}
\includegraphics[width=8.5cm,angle=0]{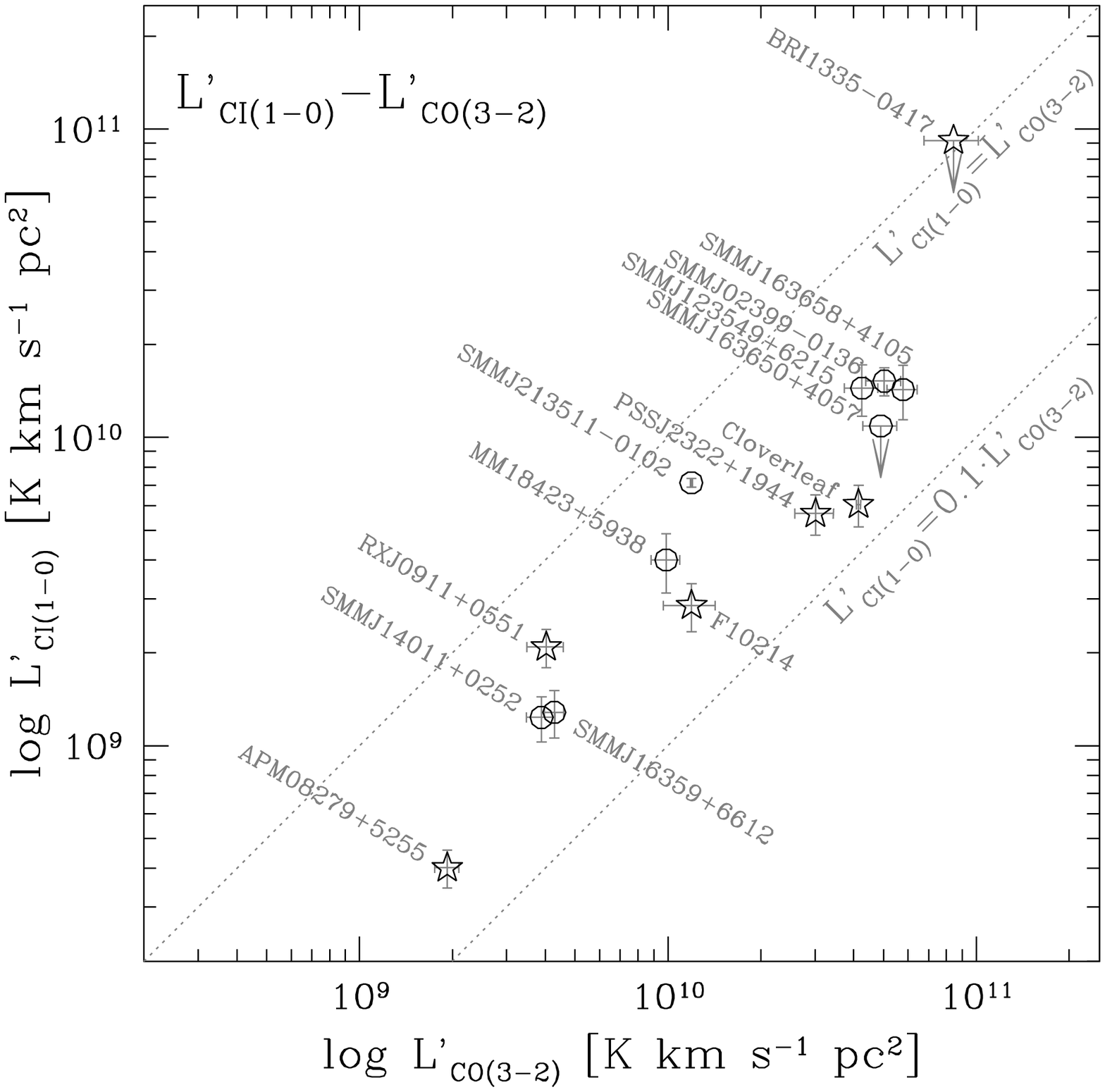}
\includegraphics[width=8.5cm,angle=0]{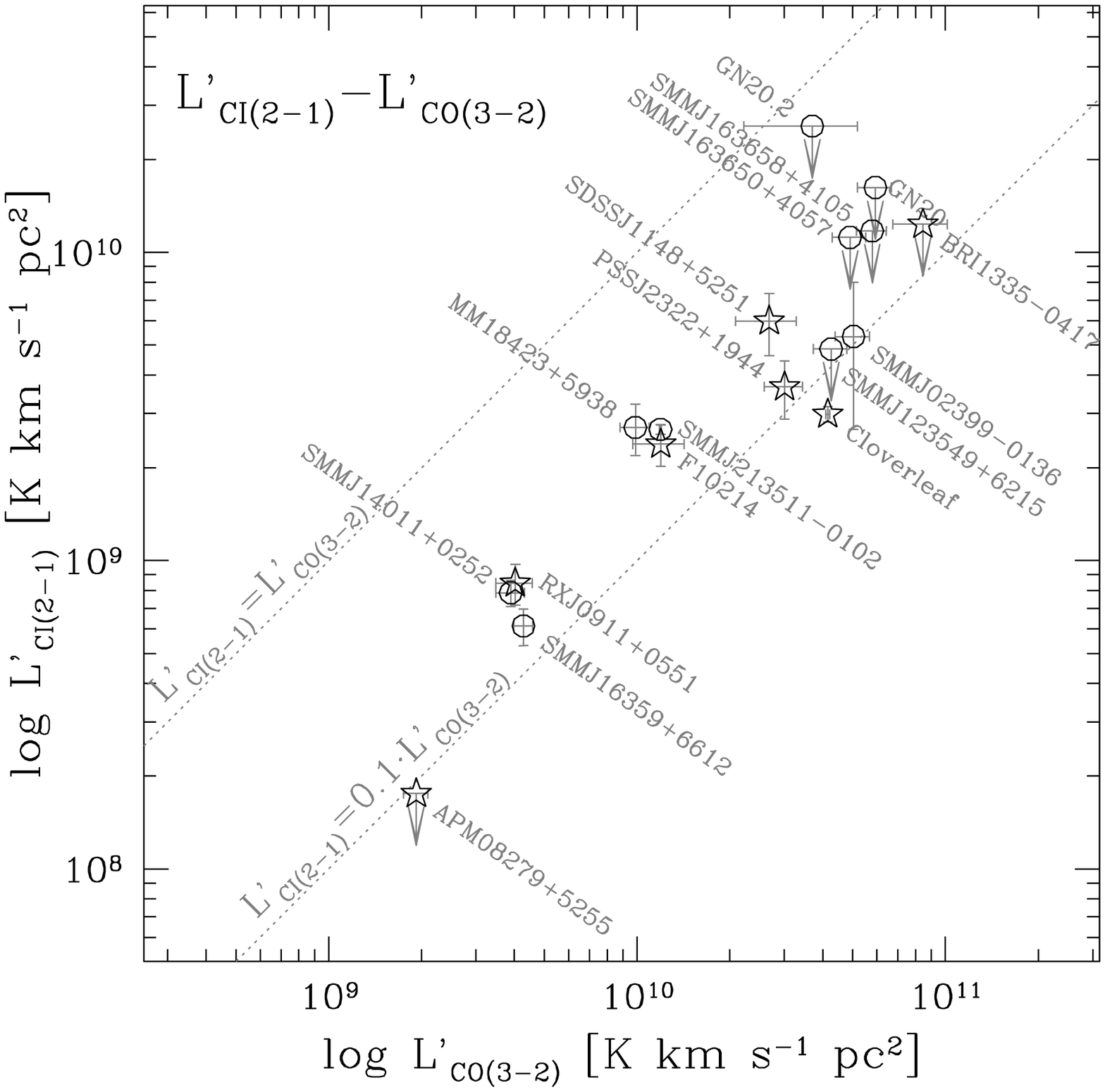}
\end{center}
\caption{$L'_\cone$ (left) and $L'_\ctwo$ (right) as a function of $L'_{\rm CO(3-2)}$ for all galaxies (stars indicate objects classified as quasars, circles: SMGs). The diagonal lines indicate $L'_\ci$=$L'_{\rm CO(3-2)}$ and $L'_\ci$=0.1$\times L'_{\rm CO(3-2)}$}
\end{figure*}

\subsection{Atomic Carbon Masses}

We now derive the atomic carbon masses of our targets. As derived in
Wei{\ss} et al.\ (2003, 2005b), the carbon mass can be calculated from
the \cone\ line luminosity via:

\begin{equation}
\label{nrmci}
M_{\ci} = 5.706\times10^{-4}\,Q(\tex)\,\frac{1}{3}\,e^{23.6/ \tex}\,L'_{\cone} [\msol] 
\end{equation}

where $Q(\tex)=1+3{\rm e}^{-T_{1}/\tex}+5{\rm e}^{-T_{2}/\tex}$ is the
\ci\, partition function. $T_{1}$\,=\,23.6\,K and $T_{2}$\,=\,62.5\,K
are the energies above the ground state\footnote{In the case of
J\,1148+5251, for which only the \ctwo\ line was observed, we use the
corresponding equation given in Wei{\ss} et al.\ 2005b}. The above
equation assumes optically thin \ci\ emission and that both carbon
lines are in LTE.

Where available we use the excitation temperature derived in
Section~4.4 above.  For those sources where only one carbon line has
been measured we assume \tex=$30$\,K. We note that the total neutral
carbon mass is not a strong function of the assumed $T_{ex}$ unless
the excitation temperature is below 20\,K (unlikely to be the case in
our targets, see also Figure~2 in Wei{\ss} et al.\ 2005a) so the exact
choice of \tex\ is not critical for the derivation of \ci\ masses.

For completeness, we also derive molecular gas (H$_2$) masses assuming
the standard conversion factor for starburst environments that should
be applicable for SMGs and QSOs ($\alpha_{\rm CO}$=M$_{\rm
gas}$/$L'_{\rm CO}$=0.8\,M$_\odot$(K\,km\,s$^{-1}$\,pc$^{2})^{-1}$,
Downes \& Solomon 1998). We here assume that the $J$=3 line is
slightly subthermally excited (i.e., $L'_{\rm CO(3-2)}$=0.9$\times
L'_{\rm CO(1-0)}$, e.g., Wei{\ss} et al.\ 2005a). The CO luminosities
(not corrected for magnification) are listed in Table~5, and the
corresponding masses (here corrected for the magnification factors
given in Table~4) in Table~6. We note that the molecular gas masses
of the submillimeter galaxies would be higher by $\sim$50\% if we
assumed the average $L'_{\rm CO(3-2)}$=0.6$\times L'_{\rm CO(1-0)}$
derived by Ivison et al.\ (2010a).

\subsection{Atomic Carbon Abundances}

We can now derive the atomic carbon abundance relative to \hh\ via
$X[\ci]/X[\hh]\,=\,M(\ci)/(6\,M(\hh))$. The carbon abundance derived
this way is independent of the magnification (ignoring potentially
differential magnification) and the applied cosmology.  We find a
carbon abundance of $X[\ci]/X[\hh]\approx\,8.4\pm3.5\times10^{-5}$ for all
sources (in agreement with earlier measurements by Ao et al., 2008,
Wei{\ss} et al., 2005b). The carbon abundance for the SMGs would
decrease if the 50\% higher molecular gas masses implied by Ivison et
al.\ (2010a) were used. For comparison, Frerking et al.\ (1989) have
derived an abundance of up to 2.2$\times$10$^{-5}$ for high extinction
regions in the Galactic star forming region Ophiuchus. Israel \& Baas
(2001, 2003) have estimated gas--phase abundances in nearby galaxies
of order 4$\times$10$^{-4}$ (their Table 6).  
We conclude that the cold molecular gas (traced by
\ci\ and CO) is already substantially enriched in our sample galaxies
at high redshift.

\begin{figure*}
\begin{center}
\includegraphics[width=8.5cm,angle=0]{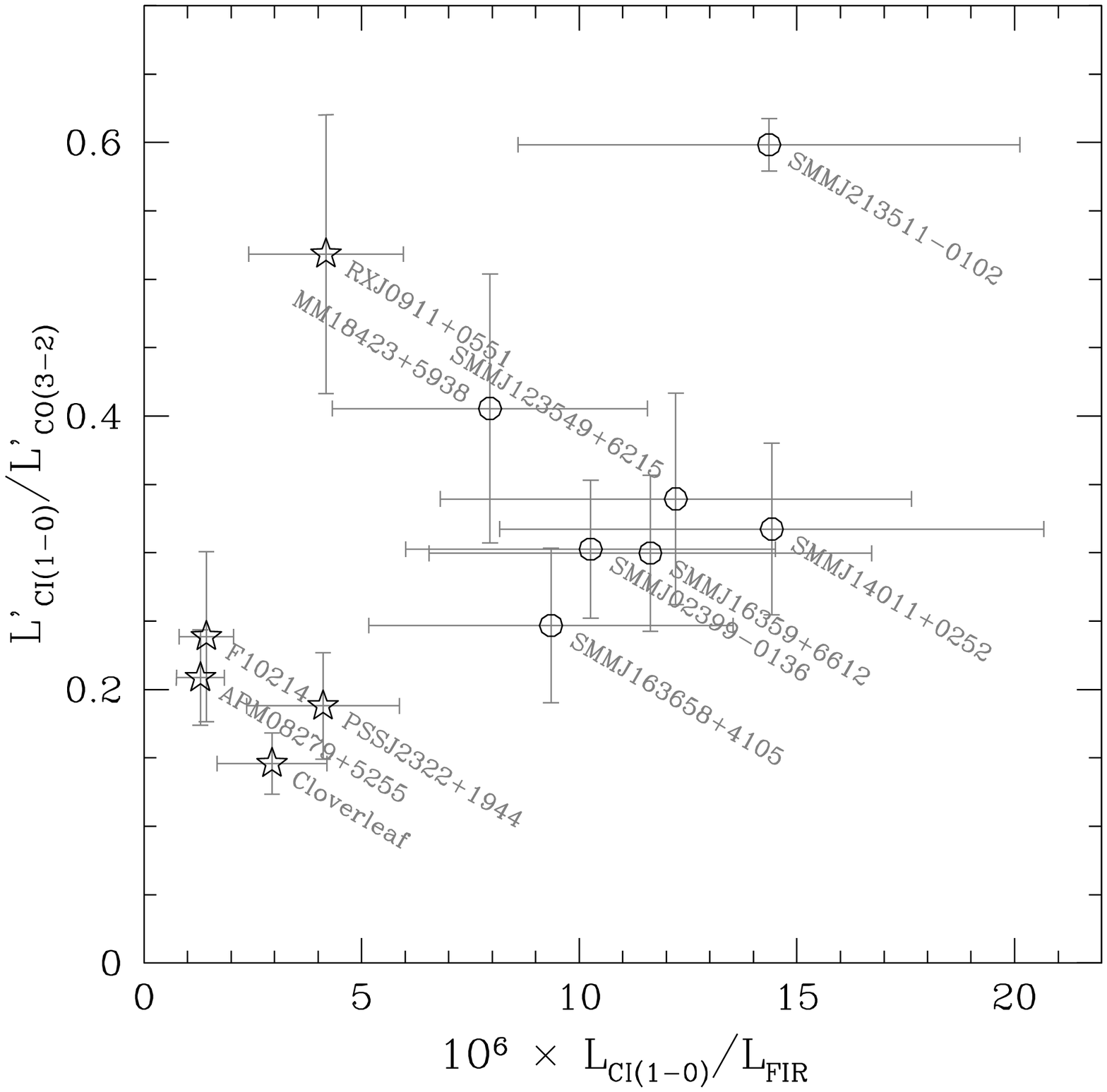}
\includegraphics[width=8.5cm,angle=0]{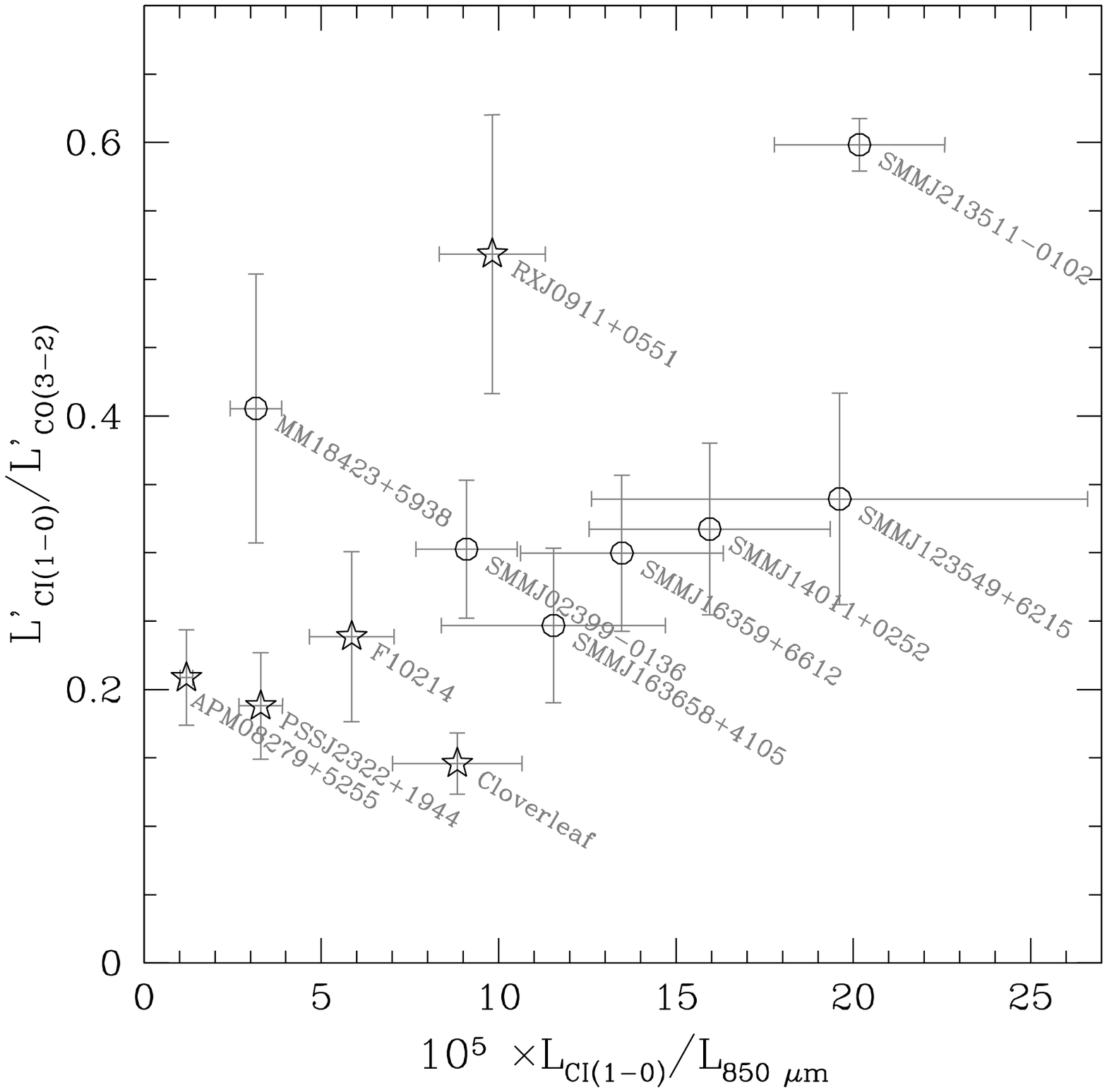}
\end{center}
\caption{{\em Left:} $L'_\cone$/$L'_{\rm CO(3-2)}$ ratio as
a function of the cooling contribution of \ci\ ($L_{{\rm \ci}}/L_{\rm
FIR}$). {\em Right:} Same $L'_\cone$/$L'_{\rm CO(3-2)}$ ratio, this time plotted as a function of 
$L_{{\rm \ci}}/L_{\rm 850\mu m}$ to take out the FIR luminosity dependence on the assumed dust temperatures. SMGs are plotted as circles, QSOs as stars in both panels. }
\end{figure*}

\begin{figure}
\begin{center}
\includegraphics[width=8.5cm,angle=0]{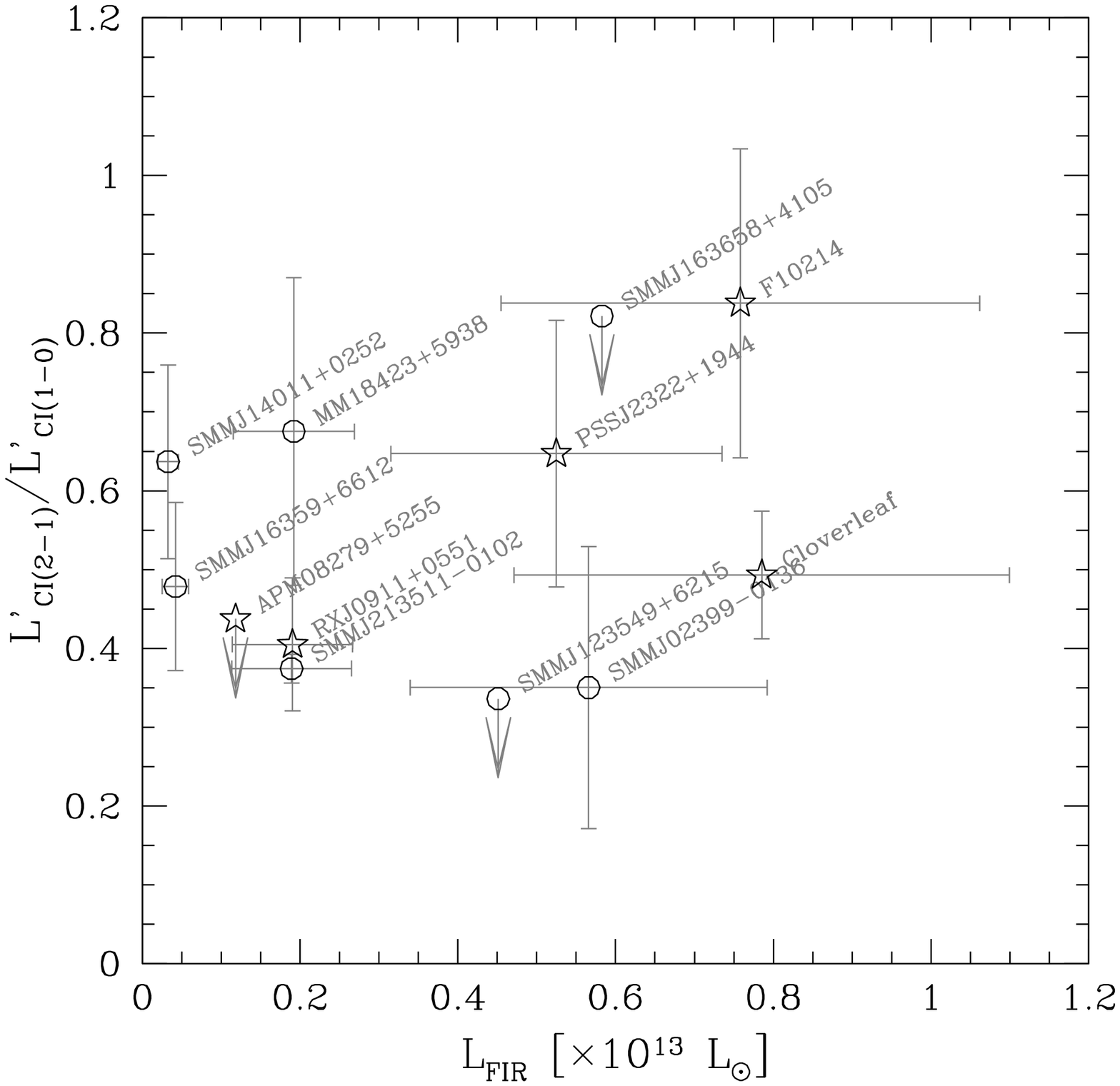}
\end{center}
\caption{$L'_\ctwo$/$L'_\cone$ ratio as a function of FIR luminosity (corrected for lensing).}
\end{figure}

\subsection{Cooling Contribution of Atomic Carbon}

As noted by Pety \etal\ (2004), a comparison of the \ci\ and FIR
luminosities shows that the cooling by \ci\ is negligible compared to
the cooling by the dust continuum (average $L_{{\rm \ci}}/L_{\rm
FIR}\sim10^{-5}$).  In Figure~14 (left) we plot the
$L'_\cone$/$L'_{\rm CO(3-2)}$ ratio as a function of the cooling
contribution of \ci\ ($L_{{\rm \ci}}/L_{\rm FIR}$). We here chose to
plot luminosity ratios to eliminate any uncertainties in magnification
corrections (ignoring the possibility of differential
magnification). Sources referred to as SMGs appear to have higher
$L'_\cone$/$L'_{\rm FIR}$ ratios than QSOs. However, part of this
offset is due to our choice of dust temperatures (Section~4.2.2): the
lower T$_{\rm dust}$ in the SMGs (compared to the QSO sample) leads to
lower $L_{\rm FIR}$ for a given S$_{\rm 850 \mu m}$ flux density,
which in turn results in higher $L_\cone$/$L_{\rm FIR}$ ratios. To
stick more to observables, we also plot the ratio of $L_{{\rm
\ci}}/L_{\rm 850\mu m}$ in the right panel of Figure~14 ($L_{\rm
850\mu m}$=4$\pi$D$_{\rm L}^2\nu$S$_{\rm 850\mu m}$;
$\nu$=352\,GHz). This is justified because for a given $L_{\rm FIR}$,
$L_{\rm 850\mu m}$ is, to first order, independent of redshift for
z$>$2 (e.g. Blain et al.\ 2002). We still find an offset (higher
$L_{{\rm \ci}}/L_{\rm 850\mu m}$) for the SMGs. This can
be partly explained by a significant contribution of the central AGN
to $L_{\rm FIR}$ in the QSOs (e.g., discussion in Wei{\ss} et al.,
2007a, for APM\,08279+5255, the source with the lowest ratio in our
study). An alternative explanation is that the cooling contribution of
\ci\ in the SMGs may be intrinsically higher than in the QSOs.

\subsection{\ci\ Ratios in the Context of PDR Models}

The PDR models of Gerin \& Phillips (2000) and our average ratio
$L_{\rm CI}$/$L_{\rm FIR}\sim10^{-5}$ suggest an
average interstellar radiation field of order 10$^3\times$G$_0$ (where
G$_0$ is the local Galactic interstellar ultraviolet radiation
field). Such high radiation fields have also been derived for other
high--z sources (e.g., Maiolino et al.\ 2005, Ivison et al.\ 2010b).

Our derived average \cone/\ctwo\ line ratio (in units of $L'$) is
0.55$\pm$0.15 which translates to a ratio of 2.7$\pm$0.9 in units of
line intensity (which scales as $\nu^3$). For a 10$^3\times$G$_0$
radiation field, the PDR models by Kaufman et al.\ (1999, their figure
8) and Meijerink et al.\ (2007, their figure 3) imply low H$_2$ volume
densities of order 10$^3$\,cm$^{-3}$. This density is of similar order
as the critical density of both \ci\ transitions which suggests that
our derived excitation temperatures are, to first order, similar to
the kinetic temperatures.  Under this assumption, the average
temperature of the \ci\ lies in between the temperatures typically
found in quiescent galactic regions (`Cirrus--like'), 15--20\,K, and
those found in active starforming environments (50--60\,K). This could
imply that the interstellar medium traced by our \ci\ observations is
a mixture of a `cold' and a `warm' molecular gas component. We note
however that the excitation temperture derived in equation~3 could be
lower than the kinetic temperature if the \ci\ excitation is not close
to LTE.

The \ci\ line ratios are not in agreement with the predictions by XDR
models of Meijerink et al.\ (2007) which suggests that gas that is
dominantly heated by an AGN should have a higher \ctwo/\cone\ ratio
than observed ($L'_\ctwo$/$L'_\cone>$0.8, Fig.~3 in Meijerink). As
shown in Figure~15 the \ci\ ratio appears to be independent of the FIR
luminosity and does not depend on the source type. Even in the case of
APM\,08279+5255, where the gas excitation is thought to be dominated
by an AGN (Wei{\ss} et al.\ 2007), the \ctwo/\cone\ ratio is lower than
predicted for an XDR environment.

\section{Discussion and Conclusions}

We have presented new \ci\ observations of 10 high--redshift
galaxies. All systems that have been looked at in \ci\ emission to
date (including studies in the literature) are very bright emitters
both in the submillimeter regime and in CO line emission, and many of
them are lensed. Even so, individual detections of the \ci\ lines have
proven to be challenging with current instrumentation. In total, we
have targeted 16 \ci\ lines in this study, and have detected 10 of
them. Including these new data, there are now a total of 17 sources in
the literature in which \ci\ has been observed (4 of which are not
detected in neither the \cone\ nor \ctwo\ emission).  

Our main finding is that the \ci\ properties of high--redshift systems
do not differ significantly from what is found in low--redshift
systems, including the Milky Way. In addition, there are no major
differences in \ci\ properties between the QSO-- and SMG--selected
samples.  We find that the $L'_\cone$/$L'_{\rm CO}$ ratios
(0.29$\pm$0.12) are similar to low--z galaxies (e.g., $0.2\pm0.2$,
Gerin \& Phillips 2000).

Measurements of both \ci\ lines offer the unique opportunity to
constrain gas excitation tempertures of the molecular gas, independent
of radiate transfer modeling. We derive an average carbon excitation
temperature of 29.1$\pm$6.3\,K for our sample. This temperature is
lower than what is typically found in starforming regions in the local
universe despite the fact that the sample galaxies have star formation
rate surface densities on kpc scales of 100's of
M$_\odot$\,yr$^{-1}$\,kpc$^{-2}$ (comparable to the most extreme
starbursts in the local universe, but on larger spatial
scales). However, the temperatures are roughly consistent with
published dust temperatures of high--redshift starforming galaxies
(Beelen et al.\ 2006, Kov{\'a}cs et al., 2006, 2010). Low carbon
excitation as well as dust temperatures could indicate that the
measurements include a significant amount of gas/dust unaffected by
star formation. However one should also keep in mind that the
conversion of excitation temperature to kinetic temperature may be
complicated by non--LTE excitation of atomic carbon.

The \ci\ abundances in our sample galaxies
(X[\ci]/X[H$_2$]=(8.4$\pm3.5)\times$10$^{-5}$) are comparable,
within the uncertainties, to what is found in local starforming
environments (Section~4.6).  This implies that the high--z galaxies
studied here are significantly enriched in carbon on galactic scales,
even though the look--back times are considerable (our average
redshift of z$\sim$3 corresponds to an age of the universe of
$\sim$2\,Gyr).

In terms of the cooling budget the \ci\ lines are a negligible coolant
(average $L_\ci$/$L_{\rm FIR}=(7.7\pm4.6)\times10^{-6}$).  We find
tentative evidence that this ratio may be elevated in the SMGs by a
factor of a few compared to the QSOs, but a larger sample will be
needed to beat down the current low--number statistics.  The increase
in sensitivity afforded by ALMA will be critical to increase the
sample size and push \ci\ studies at high redshift beyond the
brightest systems that are accessible today.

\acknowledgements We thank the referee for comments that helped to
improve the paper and thank Rowin Meijerink for useful
discussions. This work is based on observations with the IRAM Plateau
de Bure Interferometer.  IRAM is supported by INSU/CNRS (France), MPG
(Germany) and IGN (Spain).

\end{document}